\documentclass[aps,preprint,amsmath,amssymb,11pt]{revtex4}
\usepackage{graphicx}
\usepackage{epstopdf}
\usepackage[T1]{fontenc}
\usepackage{mathrsfs}
\usepackage{epsfig}
\usepackage{slashed}
\usepackage{amssymb}
\usepackage{amsmath}
\usepackage{amsthm}
\usepackage{multirow}
\usepackage{color,xcolor}

\newcommand{\pom}{\tt I\! P}

\begin{document}

\title{Probing the diffractive production of Z boson pair at forward rapidities at the LHC}
\author{Hao Sun\footnote{Corresponding author: haosun@mail.ustc.edu.cn \hspace{0.2cm} haosun@dlut.edu.cn}}
%\email[]{Corresponding author: haosun@mail.ustc.edu.cn  haosun@dlut.edu.cn}
\affiliation{
Institute of Theoretical Physics, School of Physics,
Dalian University of Technology, No.2 Linggong Road, Dalian, Liaoning, 116024, P.R.China }

\begin{abstract}
In this paper, we present the results from phenomenological analysis of Z boson pair
hard diffractive production at the LHC. The calculation is based on the Regge factorization approach.
Diffractive parton density functions extracted by the H1 Collaboration at DESY-HERA are used.
The multiple Pomeron exchange corrections are considered through the rapidity gap survival probability factor.
We give numerical predictions for single diffractive as well as double Pomeron exchange cross sections
and compare with the photon-induced and non-diffractive ones.
The contributions from quark-anti-quark collision and gluon-gluon fusion are displayed.
Various kinematical distributions are presented. We make predictions which could be compared to
future measurements at the LHC where forward proton detectors are installed and detector acceptances are considered.

\vspace{0.5cm}
% PACS numbers: 12.38.-t, 12.38.Bx, 14.70.Hp
\end{abstract}

\maketitle

\section{Introduction}

Hadronic processes can be classified as being either soft or hard,
where soft (hard) means strong interaction processes with a small (large) momentum transfer.
The hard sector is well described by perturbative Quantum Chromodynamics (pQCD)
where the coupling constant ($\rm \alpha_s$) is small (compare to the 'hard' momentum transfer)
and a perturbative expansion in terms proportional to powers of $\rm \alpha_s$ works.
This is done by means of QCD factorization\cite{QCDfactoraztion1}\cite{QCDfactoraztion2}
which has been thoroughly tested and taken as the most powerful tool in describing high energy hadronic collisions.
On the other hand, soft processes are characterised by an energy scale of the order of the hadron size ($\rm 1\ fm \simeq 200\ MeV$)
where $\rm \alpha_s$ is large enough to make the higher order terms non-negligible,
thus making the soft processes intrinsically non-perturbative.
To gain understanding of soft or non-perturbative QCD, it is therefore advantageous to first
consider soft effects in hard scattering events, since the hard scale gives a firm ground in
terms of a parton level process which is calculable in pQCD. This hard-soft interplay is the
basis for the research field of diffractive hard scattering.

Encoding the parton distribution functions (PDFs),
one can separate the hard perturbation contributions from the soft non-perturbative ones.
Following this idea, factorization is still being used and has been carefully proved
in diffractive Deep Inelastic Scattering (DDIS)\cite{factorizationProof}.
In the framework of Regge factorization, the so called Ingelman and Schlein (IS) model\cite{Ingelman-Schlein}
has been largely used in describing hard diffractive events in electron-proton (ep) collisions\cite{diffractive_HERA}.
The IS model essentially considers that diffractive scattering is attributed to the exchange of a Pomeron,
i.e. a colorless object with vacuum quantum numbers.
The Pomeron is treated like a real particle, and one considers that a diffractive ep collision is due
to an electron-Pomeron collision and that a diffractive proton-proton (pp) collision is due to a proton-Pomeron collision.
However, the nature of the Pomeron and its reaction mechanisms are still unknown.
Diffractive study may help us understanding more about the QCD Pomeron structure.
One should be careful that factorization seems to be broken when going from DDIS at
HERA to hadron-hadron collisions at the Tevatron and the Large hadron collider (LHC).
Theoretical studies\cite{Theory_Multiple_Scatter} predicted that the breakdown of the factorization
is due to soft rescattering corrections associated to reinteractions
(referred to as multiple scatterings effects) between spectator partons of the colliding hadrons
that fill in the rapidity gaps related to Pomeron exchange.

In order to constrain the modelling of the gap survival effects and also improve our limited
understanding of diffraction, it will be crucial to, in experimental point of view,
discriminate the diffractive production from the non-diffractive processes.
Indeed, diffractive events can be characterized by having a rapidity gap (RG), say, a region in rapidity
or polar angle without any particles. Another definition is to require a leading particle
carrying most of the beam particle momentum, which is kinematically related to
a RG. These RGs in the forward or backward rapidity regions, connect directly to
the soft part of the events, and therefore non-perturbative effects, on a long space-time scale.
Thus, the experimental signature for diffractive production
is either the presence of one(two) RG(s) in the detector or one(both) proton(s) tagged in the final state(s).
The potential for using RG vetoes to select diffractive events are highly favoured
by the newly installed HERSCHEL forward detectors\cite{HERSCHEL_forward_detector} at LHCb,
due to its low instantaneous luminosity and wide rapidity coverage.
Similar scintillation counters are also installed at ALICE\cite{ALICE_forward_detector} and CMS\cite{HERSCHEL_forward_detector}.
Potentially intact proton(s) tagging to select(or exclude) exclusive(or diffractive) events
can be realized by using the approved AFP\cite{AFP_detector} and installed CT-PPS\cite{CT-PPS_detector}
forward proton spectrometers, associated with the ATLAS and CMS central detectors\cite{Royon_forward_detector} at the LHC.
The installation of forward detectors at the LHC may provide possibility, somehow open a new window to study new physics
at TeV scale, whereas diffractive events may serve as one of the most important background source.
Besides Regge factorization or the amount of gap survival probability which are widely accepted approximations,
resonance production, in the central and forward (proton excitation) regions, is also an important issue.
Related studies can be found, i.e., in refs\cite{Resonance_Jenkovszky}\cite{Resonance_Jenkovszky1}\cite{Resonance_Fiore}.
In any case, diffractive productions worth being carefully studied and precisely estimated.

A lot of works on diffraction can be found in the literatures for a long time which include, i.e.,
diffractive dijet\cite{Ingelman-Schlein},
heavy flavour jets\cite{SD_DD_jj_Machado}\cite{SD_DD_heavyJ_Ducati}\cite{SD_DD_heavyJ_Heyssler},
Drell-Yan pair\cite{SD_DY_LHC_Ceccopieri}, photon\cite{SD_DD_rr_LHC_Goncalves}
and also diffractive Higgs productions\cite{Diff_Higgs_Heyssler}\cite{SD_Higgs_Ducati}\cite{SD_Higgs_Enberg}\cite{SD_Higgs_Erhan}, etc.
In our present paper, we concentrate on the hard diffractive Z boson pair production at the LHC.
Diffractive hadroproduction of single electroweak boson
was first observed experimentally at the Tevatron\cite{Diff_CDF_Bosons}.
Theoretical analysis were presented in
\cite{SD_DD_WZ_Bruni}\cite{SD_DD_WZ_Alvero}\cite{Diff_Tevatron_WsJets_Covolan}\cite{Diff_Tevatron_WsJets_Covolan1}
at the Tevatron, in \cite{Diff_RIHC_rWZ} at the RHIC,
and in \cite{Diff_RIHC_rWZ}\cite{Diff_LHC_WZ}\cite{Diff_LHC_W}\cite{DIff_LHC_WZ1}\cite{Diff_LHC_W_Ingelman} at the LHC.
Typically, ref.\cite{Diff_LHC_W} show that single diffractive W boson production
asymmetry in rapidity is a particularly good observable
at the LHC to test the concept of the flavour symmetric Pomeron parton distributions
and may provide an additional constraint for the PDFs in the proton.
Ref.\cite{DIff_LHC_WZ1} show that diffractive gauge bosons production can be useful to constrain the modelling of the gap survival effects.
All these referees show that by using gauge boson productions, studies of the Pomeron structure and diffraction phenomenology are feasible.
For diboson production, diffractive W boson pair is the frontier one which have been studied in refs.\cite{WW_diff_Royon}\cite{WW_diff_Lebiedowicz}.
The Z boson pair diffractive is less important due to its small production rate compare to W boson pair production.
Nevertheless, at the LHC high energy frontier, still worth being studied rather than at the Tevatron.

Our paper is organized as follows:
in section 2 we present the production mechanisms starting from general production to diffractive ones.
We show the details concerning the parameterization for the diffractive PDFs in the Pomeron.
In addition, we present the theoretical estimations for the gap survival probability factor.
Typically, the forward detector acceptances are considered. We present our numerical results
and perform predictions to future measurements at the LHC in section 3.
Finally we make our summary in the last section.

\section{CALCULATION FRAMEWORK}

\subsection{Production Mechanism}

\begin{figure}[hbt!]
\centering
\includegraphics[scale=0.30]{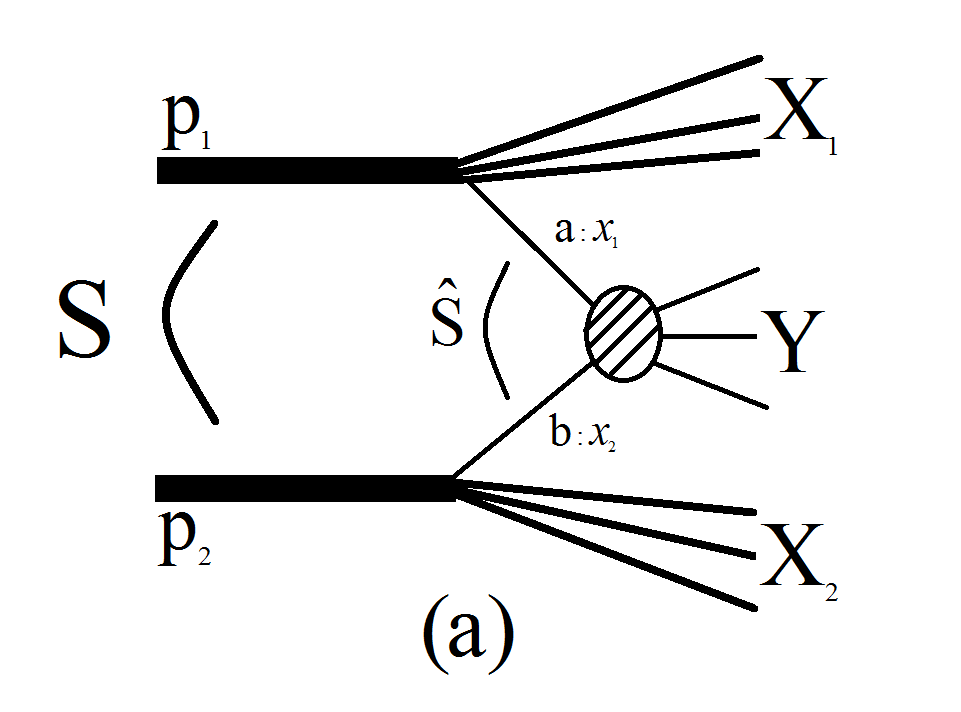}
\includegraphics[scale=0.30]{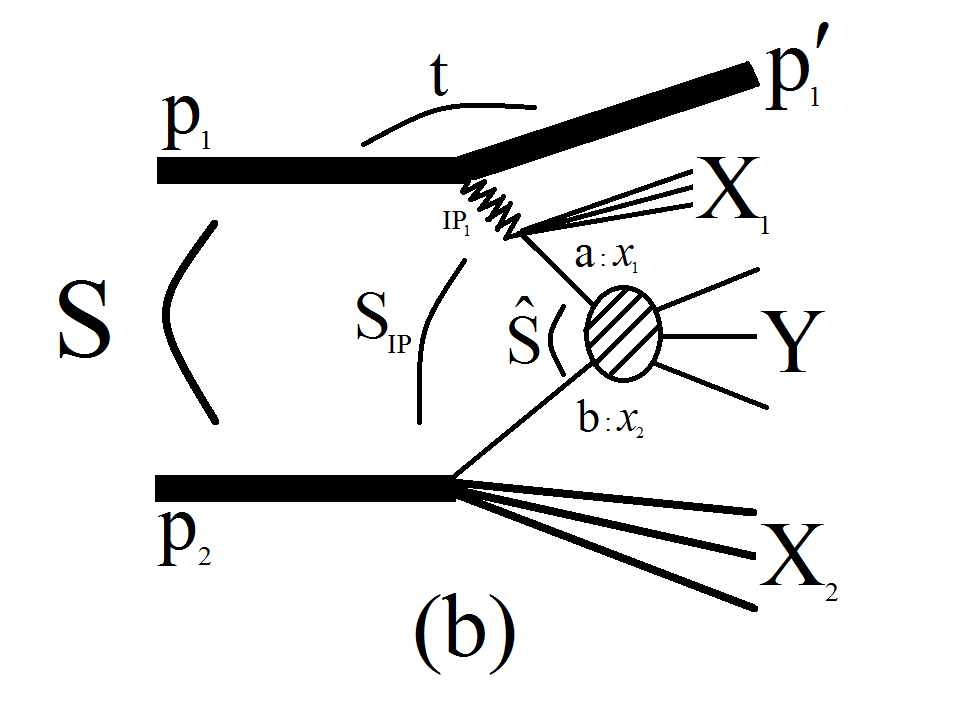}
\includegraphics[scale=0.30]{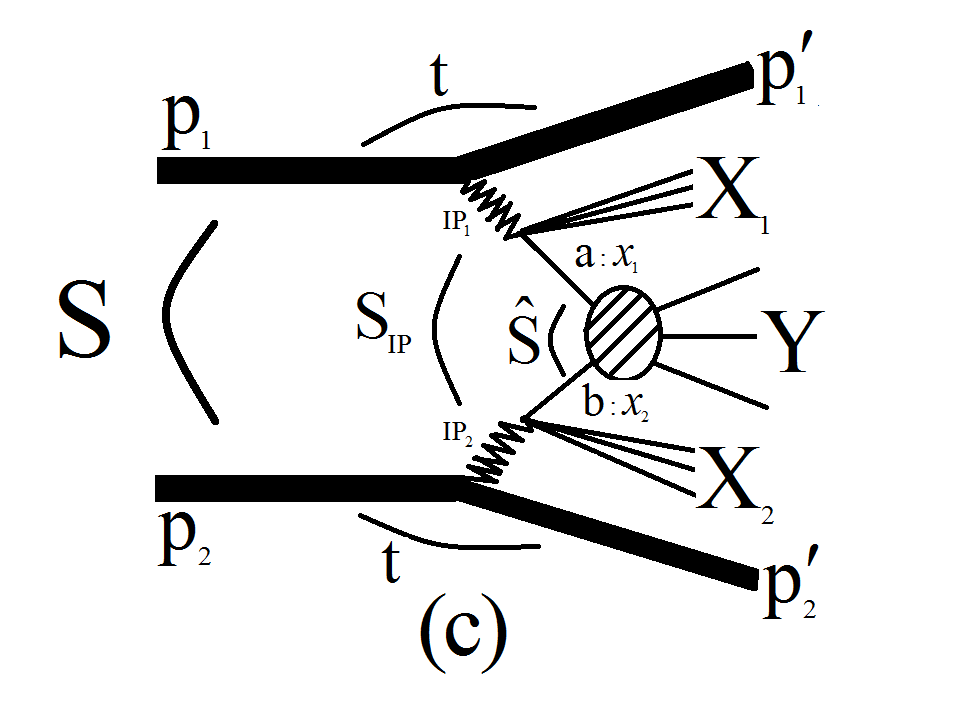}
\caption{\label{fig1abc}
Illustrated diagrams for the non-diffractive (a), single diffractive (b) and double Permon exchange (c) production.}
\end{figure}

Our starting point is the introduction of the general inclusive total cross section for the process
\begin{eqnarray}
\rm p_1 + p_2 \to (a+b\to Y) + X
\end{eqnarray}
in Fig.\ref{fig1abc}(a), in which partons of two hadrons (a from $\rm p_1$ and b from $\rm p_2$)
interact to produce a Y system, at the center of mass (CMS) energy $\rm \sqrt{s}$.
The total hadronic inclusive cross section is obtained by convoluting the total partonic cross section
with the PDFs of the initial hadrons,
\begin{eqnarray} \nonumber
\rm \sigma_{p_1p_2\to Y+X} (s, \mu^2_F, \mu^2_R) &=& \rm \sum_{a,b=q,\bar{q},g}
    \int^1_{\tau_0}\ 2zdz \int^1_{z^2}\ \frac{dx_1}{x_1} \ f_{a/p_1}(x_1,\mu^2_F) \ f_{b/p_2}( \frac{z^2}{x_1},\mu^2_F) \\
\rm & &\rm  \hat\sigma_{a+b\to Y}(\hat{s}=z^2 s,\ \mu^2_F,\ \mu^2_R) + (a \rightleftharpoons b)
\label{InclusiveXsec}
\end{eqnarray}
where the sum $\rm a,b=q,\bar{q},g$ is over all massless partons.
$\rm z^2=x_1x_2$ with $\rm x_1$ and $\rm x_2$ are the hadron momentum fractions carried by the interacting partons.
The partonic cross section is $\rm \hat\sigma_{a+b\to Y}(\hat{s}, \mu^2_F, \mu^2_R)$ where
$\rm \hat{s}$ is the partonic CMS energy, $\rm \mu_F(\mu_R)$ is the renormalization(factorization) scale.
$\rm \tau_0=m_{Y}/\sqrt{s}$ and $\rm m_{Y}$ is the mass threshold for Y system.
$\rm f_{i/p}(x_i,\mu^2_F)$ is the PDF of a parton of flavour i in the hadron p,
and are evaluated at the factorization scale (usually assumed to be equal to the renormalization scale).

For the hard diffractive processes, we will consider the Ingelman-Schlein (IS) picture\cite{Ingelman-Schlein},
in which a Pomeron structure (with quark and gluon content) is introduced.
In the expression for single diffractive (SD) processes, there includes three steps:
First, one of the hadrons, say hadron $\rm p_1$ with energy E, emits a Pomeron($\pom_1$),
with only a small squared four momentum transfer $\rm |t|$,
and turns to hadron $\rm p'_1$ with energy $\rm E'$ but remains almost intact.
Second, the remaining hadron scatters off the emitted Pomeron.
Partons from the Pomeron interact with partons from the other hadron($\rm p_2$) and produce a Y system.
Finally, hadron $\rm p'_1$ is detected in the final state with a reduced energy loss (defined as $\rm \xi=(E-E')/E$) by proposed
forward proton detectors\cite{AFP_detector}\cite{CT-PPS_detector}.
Meantime, the Y system and the remaining remnants(X) go to the general central detectors.
A typical SD reaction is presented in Fig.\ref{fig1abc}(b) and can be given as
\begin{eqnarray}
\rm p_1 + p_2 \to p_1 + (a + b \to Y) + X.
\end{eqnarray}
In the IS approach, the SD cross section is assumed to factorize into the total
Pomeron-hadron cross section and a Pomeron flux factor\cite{Ingelman-Schlein}.
This means we can replace the PDFs in Eq.(\ref{InclusiveXsec}) by
\begin{eqnarray} \nonumber
\rm x_{i} f_{i/p}(x_{i},\mu^2) \ \Rightarrow \
x_{i} f^D_{i/p}(x_{i},\mu^2) &=&\rm \int dx_{\pom} \int d\beta \ \bar{f}(x_{\pom}) \cdot \beta f_{i/\pom}(\beta,\mu^2) \cdot \delta(\beta-\frac{x_i}{x_{\pom}}) \\
&\equiv& \rm \int dx_{\pom} \ \bar{f}(x_{\pom}) \frac{x_i}{x_{\pom}} \ f_{i/\pom}(\frac{x_i}{x_{\pom}},\mu^2)
\end{eqnarray}
with the defined quantity $\rm \bar{f}(x_{\pom}) \equiv \int^{t_{max}}_{t_{min}} f_{\pom/p}(x_{\pom},t) dt$.
Here $\rm \beta f_{i/\pom}(\beta,\mu^2)$ is the PDF of a parton of flavour i in the Pomeron and
$\rm f_{\pom/p}(x_{\pom},t)$ is the Pomeron flux factor, describe the emission rate of Pomerons by the hadron.
$\rm x_{\pom}$ is the Pomeron kinematical variable defined as $\rm x_{\pom}=s_{\pom_1p_2}/s_{p_1p_2}$, where
$\sqrt{s_{\pom_1 p_2}}$ is the CMS energy in the Pomeron-hadron system and $\rm \sqrt{s_{p_1p_2}}\equiv\sqrt{s}$ is the CMS energy
in the hadron($\rm p_1$) hadron($\rm p_2$) system.
The single diffractive cross section can be written as
\begin{eqnarray}\nonumber
&&\rm \sigma^{SD}_{p_1p_2\to p_1+Y+X} (s, \mu^2_F, \mu^2_R) \\\nonumber
&=& \rm \sum_{a,b=q,\bar{q},g}
    \int^1_{\tau_0} 2zdz \int^1_{z^2} \frac{dx_1}{x_1} \ f^D_{a/p_1}(x_1,\mu^2_F) f_{b/p_2}(\frac{z^2}{x_1},\mu^2_F) \\ \nonumber
&&\rm  \hat\sigma_{a+b\to Y}(\hat{s}=z^2 s, \mu^2_F,\ \mu^2_R) + (a \rightleftharpoons b)  \\\nonumber
&=& \rm \sum_{a,b=q,\bar{q},g}
    \int^1_{\tau_0}\ 2zdz \int^1_{z^2} \frac{dx_1}{x_1} \int^{x^{max}_{\pom}}_{x_1} \frac{dx_{\pom}}{x_{\pom}}
    \bar{f}_{\pom/p_1}(x_{\pom})  \ f_{a/\pom}(\frac{x_1}{x_{\pom}},\mu^2_F)
    \ f_{b/p_2}( \frac{z^2}{x_1},\mu^2_F)  \\
&&\rm    \hat\sigma_{a+b\to Y}(\hat{s}=z^2 s,\ \mu^2_F,\ \mu^2_R) + (a \rightleftharpoons b).
\label{SDXsec}
\end{eqnarray}

A similar factorization can also be applied to double Pomeron exchange (DPE) process,
where both colliding hadrons can be detected in the final states.
This diffractive process is also known as central diffraction (CD) production.
The illustration diagram is presented in Fig.\ref{fig1abc}(c). A typical DPE reaction is given as
\begin{eqnarray}
\rm p_1 + p_2 \to p_1 + (a + b \to Y) + X + p_2.
\end{eqnarray}
The total cross section for DPE processes reads as
\begin{eqnarray}\nonumber
&&\rm \sigma^{DPE}_{p_1p_2\to p_1+Y+X+p_2} (s, \mu^2_F, \mu^2_R) \\\nonumber
&=& \rm \sum_{a,b=q,\bar{q},g}
    \int^1_{\tau_0} 2zdz \int^1_{z^2} \frac{dx_1}{x_1} \ f^D_{a/p_1}(x_1,\mu^2_F) f^D_{b/p_2}(\frac{z^2}{x_1},\mu^2_F)
    \hat\sigma_{a+b\to Y}(\hat{s}=z^2 s, \mu^2_F,\ \mu^2_R)  \\\nonumber
&=& \rm \sum_{a,b=q,\bar{q},g}
    \int^1_{\tau_0}\ 2zdz \int^1_{z^2} \frac{dx_1}{x_1}
    \int^{x^{max}_{\pom_1}}_{x_1} \frac{dx_{\pom_1}}{x_{\pom_1}}
    \bar{f}_{\pom_1/p_1}(x_{\pom_1})  \ f_{a/\pom_1}(\frac{x_1}{x_{\pom_1}},\mu^2_F) \\
&&\rm  \int^{x^{max}_{\pom_2}}_{z^2/x_1} \frac{dx_{\pom_2}}{x_{\pom_2}}
       \bar{f}_{\pom_2/p_2}(x_{\pom_2})  \ f_{b/\pom_2}(\frac{z^2}{x_1x_{\pom_2}},\mu^2_F)  \hat\sigma_{a+b\to Y}(\hat{s}=z^2 s,\ \mu^2_F,\ \mu^2_R).
\label{DPEXsec}
\end{eqnarray}

\subsection{The Pomeron Structure Function}

In order to estimate the diffractive cross sections, two elements are needed:
\begin{itemize}
\item $\rm f_{i/\pom}(x_i,\mu^2)$: the diffractive parton distribution function (dPDF)
which describe a perturbative distribution of partons in the Pomeron.
We will consider the dPDFs extracted by the H1 collaboration at DESY-HERA\cite{H1dPDF}.
\item $\rm f_{\pom/p}(x_{\pom},t)$: the Pomeron flux factor which describe the ``emission rate'' of Pomeron by the hadron and
represents the probability that a Pomeron with particular values of $\rm (x_{\pom},t)$ couples to the proton.
\end{itemize}

The dPDFs are modelled in terms of a light flavour singlet distribution $\rm \Sigma(z)$,
consisting of u, d and s quarks and anti-quarks with $\rm u=d=s=\bar{u}=\bar{d}=\bar{s}$,
and a gluon distribution g(z). Here z is the longitudinal momentum fraction of the parton
entering the hard sub-process with respect to the diffractive exchange, such that $\rm z=\beta$
for the lowest order quark-parton model process, whereas $\rm 0<\beta<z$ for higher order processes.
The quark singlet and gluon distributions are parameterised at $\rm Q^2_0$ using the general form
\begin{eqnarray}
\rm zf_{i}(z,Q^2_0)=A_i z^{B_i}(1-z)^{C_i}exp[-\frac{0.01}{1-z}],
\label{generalForm}
\end{eqnarray}
where the last exponential factor ensures that the dPDF's vanish at z=1,
as required for the evolution equations to be solvable.
For the quark singlet distribution, the data require the inclusion of all three parameters
$\rm A_q$ , $\rm B_q$ and $\rm C_q$ in Eq.(\ref{generalForm}).
By comparison, the gluon density is weakly constrained by the data, which is found
to be insensitive to the $\rm B_g$ parameter. The gluon density is thus parameterized at $\rm Q^2_0$
using only the $\rm A_g$ and $\rm C_g$ parameters. With this parameterization,
one has the value $\rm Q^2_0=1.75\ GeV^2$ and it is referred to as the ``H1 2006 dPDF Fit A''.
It is verified that the fit procedure is not sensitive to the gluon PDF and a new adjust was done
with $\rm C_g=0$. Thus, the gluon density is then a simple constant at the starting scale for evolution,
which was chosen to be $\rm Q^2_0=2.5\ GeV^2$ and it is referred to as the ``H1 2006 dPDF Fit B''.

For the Pomeron flux factor, we apply the standard flux form from Regge phenomenology \cite{ReggeTheory},
based on the Donnachie-Landshoff model \cite{Donnachie-Landshoff-model}\cite{Donnachie-Landshoff-model-2}.
The $\rm x_{\pom}$ dependence is parameterised by
\begin{eqnarray}
\rm f_{\pom/p}(x_{\pom},t) = A_{\pom} \cdot \frac{e^{B_{\pom}t}}{x^{2\alpha_{\pom}(t)-1}_{\pom}}
\end{eqnarray}
where the Pomeron Regge trajectory is assumed to be linear, $\rm \alpha_{\pom}(t)=\alpha_{\pom}(0)+\alpha'_{\pom}t$,
and the parameters $\rm B_{\pom}$ and $\rm \alpha'_{\pom}$
and their uncertainties are obtained from fits to H1 FPS data\cite{H1FPSdata}.
In our calculation, we take $\rm \alpha_{\pom}(0)=1.1182\pm0.008$ in fit A ($\rm \alpha_{\pom}(0)=1.1110\pm0.007$ in fit B),
$\rm B_{\pom}=5.5^{-2.0}_{+0.7}\ GeV^{-2}$ and $\rm \alpha'_{\pom} = 0.06^{+0.19}_{-0.06}\ GeV^{-2}$.
The value of the normalization parameter $\rm A_{\pom}$ is chosen such that
$\rm x_{\pom}\cdot\int^{t_{max}}_{t_{min}}f_{\pom/p}(x_{\pom},t)dt=1$ at $\rm x_{\pom}=0.003$,
where $\rm t_{max}\simeq - \frac{m^2_px^2_{\pom}}{1-x_{\pom}}$ is the maximum kinematically
accessible value of $\rm t$, $\rm m_p=0.93827231\ GeV$ is the proton mass and $\rm t_{min}=-1.0\ GeV^2$
is the limit of the measurement. So we get
\begin{eqnarray}
\rm A_{\pom} =
\frac{ x_{\pom}^{2\alpha_{\pom}(0) -2} (B_{\pom}-2\alpha'_{\pom}\ \text{ln} x_{\pom})}
{exp[-(B_{\pom}-2\alpha'_{\pom}\ \text{ln} x_{\pom})\frac{m^2_px^2_{\pom}}{1-x_{\pom}}]-exp[-(B_{\pom}-2\alpha'_{\pom}\ \text{ln} x_{\pom})]}
\ \text{with}\ x_{\pom}=0.003.
\end{eqnarray}
Thus we have
\begin{eqnarray} \nonumber
\rm \bar{f}(x_{\pom}) &=&\rm  \frac{A_{\pom}}{ x_{\pom}^{2\alpha_{\pom}(0) -1} (B_{\pom}-2\alpha'_{\pom}\ \text{ln} x_{\pom}) }  \\
&&\rm \cdot \left[
exp[-(B_{\pom}-2\alpha'_{\pom}\ \text{ln} x_{\pom})\frac{m^2_px^2_{\pom}}{1-x_{\pom}}]-exp[-(B_{\pom}-2\alpha'_{\pom}\ \text{ln} x_{\pom})]
\right].
\end{eqnarray}

\subsection{Multiple-Pomeron Scattering Corrections}

We have assumed Regge factorization which is known to be violated in hadron-hadron collisions.
Theoretical studies predicted that the violation is due to
the soft interactions between spectator partons of the colliding hadrons,
which lead to an extra production of particles that fill in the rapidity gaps related to Pomeron exchange.
So that when the rapidity gaps are measured, one has to include absorption effect in the formalism of
the resolved Pomeron. Different models of absorption corrections (one-, two- or three-channel approaches) for
diffractive processes were presented in the literature.
The absorption effects for the diffractive processes
were calculated e.g. in \cite{Multi_Pomeron_Correction_Cisek}\cite{Multi_Pomeron_Correction_Khoze}\cite{Multi_Pomeron_Correction_Maor}.
The different models give slightly different predictions.
Usually an average value of the gap survival probability $\rm \langle|S|^2\rangle$ is calculated
first and then the cross sections for different processes is multiplied by this value.
Here we shall follow this simplified approach.
The survival probability depends on the collision energy and can be sometimes parameterized as:
\begin{eqnarray}
\rm \langle|S|^2\rangle(\sqrt{s}) = \frac{a}{b+ln(\sqrt{s/s_0})}
\label{S2}
\end{eqnarray}
with a = 0.126, b=-4.688 and $\rm s_0=1\ GeV^2$ and more details can be found in original publications.
This formula gives typical value of survival probabilities for diffractive production in
proton-proton collisions of $4.5\%$ at Tevatron and $2.6\%$ at the LHC.
Indeed, more precise value should be updated by measurements.
For example, from the diffractive cross sections at the 8 TeV LHC
one gets typically value of $\rm \langle|S|^2\rangle=8\%$
extracted by the CMS collaboration for diffractive dijet production\cite{Svalue_jj8LHC}.
For the SD production and DPE production there should be some difference for the value of the factors.
Probable uncertainty may as large as 30 percent, which is one of the largest uncertainty source in diffractive production
and should able to be reduced thanks to the forthcoming measurements at the LHC.

\subsection{Forward Detector Acceptance}

We assume the intact protons in diffractive events to be tagged in the forward proton detectors of the
CMS-TOTEM Collaborations\cite{CT-PPS_detector}, or those to be installed by the ATLAS Collaboration
in the future called AFP detectors\cite{AFP_detector}.
The idea is to measure scattered protons at very small angles at the interaction point and to use the LHC
magnets as a spectrometer to detect and measure them. We use the following acceptances\cite{acceptance}:
\begin{itemize}
 \item $0.015<\xi<0.15$ for ATLAS-AFP
 \item $0.0001<\xi<0.17$ for TOTEM-CMS.
\end{itemize}
These acceptances correspond to cuts on longitudinal momentum fractions of outgoing protons.
To obtain the constrained diffractive PDFs, we convolute the Pomeron flux with the Pomeron PDFs while
imposing a reduction in the phase space of $\xi$. Imaging a reduced energy loss can be probed
in the range $\rm \xi_{min}<\xi<\xi_{max}$, we can write the final $\xi$ dependent SD cross section as\cite{FDA_Marquet}
\begin{eqnarray}\nonumber
&&\rm \sigma^{SD}_{p_1p_2\to p_1+Y+X} (s, \mu^2_F, \mu^2_R) \\\nonumber
&=&\rm  \langle|S|^2\rangle_{SD} \sum_{a,b=q,\bar{q},g}
    \int^1_{\tau_0}\ 2zdz \int^1_{z^2} \frac{dx_1}{x_1} \int^{Min(x^{max}_{\pom},\xi_{max})}_{Max(x_1,\xi_{min})} \frac{dx_{\pom}}{x_{\pom}}
    \bar{f}_{\pom/p_1}(x_{\pom})  \ f_{a/\pom}(\frac{x_1}{x_{\pom}},\mu^2_F)  \\
&&\rm  f_{b/p_2}( \frac{z^2}{x_1},\mu^2_F) \hat\sigma_{a+b\to Y}(\hat{s}=z^2 s,\ \mu^2_F,\ \mu^2_R) + (a \rightleftharpoons b)
\label{xiSDXsec}
\end{eqnarray}
The final cross section for the DPE processes can be written as\cite{FDA_Marquet}
\begin{eqnarray}\nonumber
&&\rm \sigma^{DPE}_{p_1p_2\to p_1+Y+X+p_2} (s, \mu^2_F, \mu^2_R) \\\nonumber
&=&\rm  \langle|S|^2\rangle_{DPE}  \sum_{a,b=q,\bar{q},g}
    \int^1_{\tau_0}\ 2zdz \int^1_{z^2} \frac{dx_1}{x_1}
    \int^{Min(x^{max}_{\pom_1},\xi_{max})}_{Max(x_1,\xi_{min})} \frac{dx_{\pom_1}}{x_{\pom_1}}
    \bar{f}_{\pom_1/p_1}(x_{\pom_1})  \ f_{a/\pom_1}(\frac{x_1}{x_{\pom_1}},\mu^2_F) \\
&&\rm  \int^{Min(x^{max}_{\pom_2},\xi_{max})}_{Max(z^2/x_1,\xi_{min})} \frac{dx_{\pom_2}}{x_{\pom_2}}
       \bar{f}_{\pom_2/p_2}(x_{\pom_2})  \ f_{b/\pom_2}(\frac{z^2}{x_1x_{\pom_2}},\mu^2_F)  \hat\sigma_{a+b\to Y}(\hat{s}=z^2 s,\ \mu^2_F,\ \mu^2_R).
\label{xiDPEXsec}
\end{eqnarray}

\section{NUMERICAL RESULTS}

At parton level, Z pair hadronic production is induced by quark-anti-quark collision mode at the leading order (LO).
For gluon-gluon (and $\gamma\gamma$ fusion for photoproduction) fusion initial state,
the LO contribution is induced at one loop level due to the missing of the tree contribution.
We perform our numerical calculations with in-house coding based on FeynArts,
FormCalc and LoopTools (FFL) package\cite{FeynArts,FormCalc,LoopTools}.
We adopt BASES\cite{BASES} to do the phase space integration.
In what follows, we present predictions for hard diffractive production of Z boson pair based on previous discussion.

\begin{figure}[hbt!]
\centering
\includegraphics[scale=0.40]{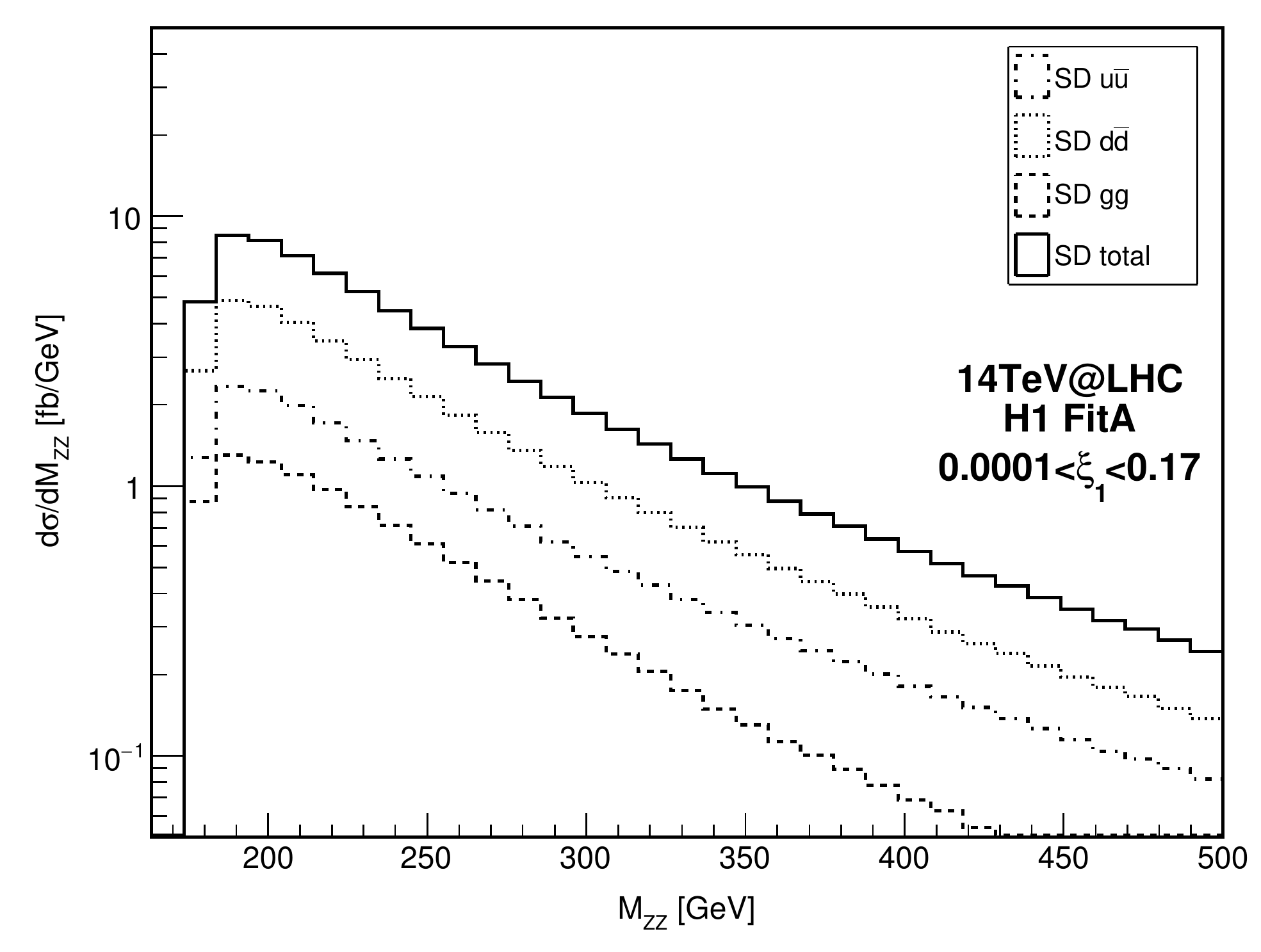}
\includegraphics[scale=0.40]{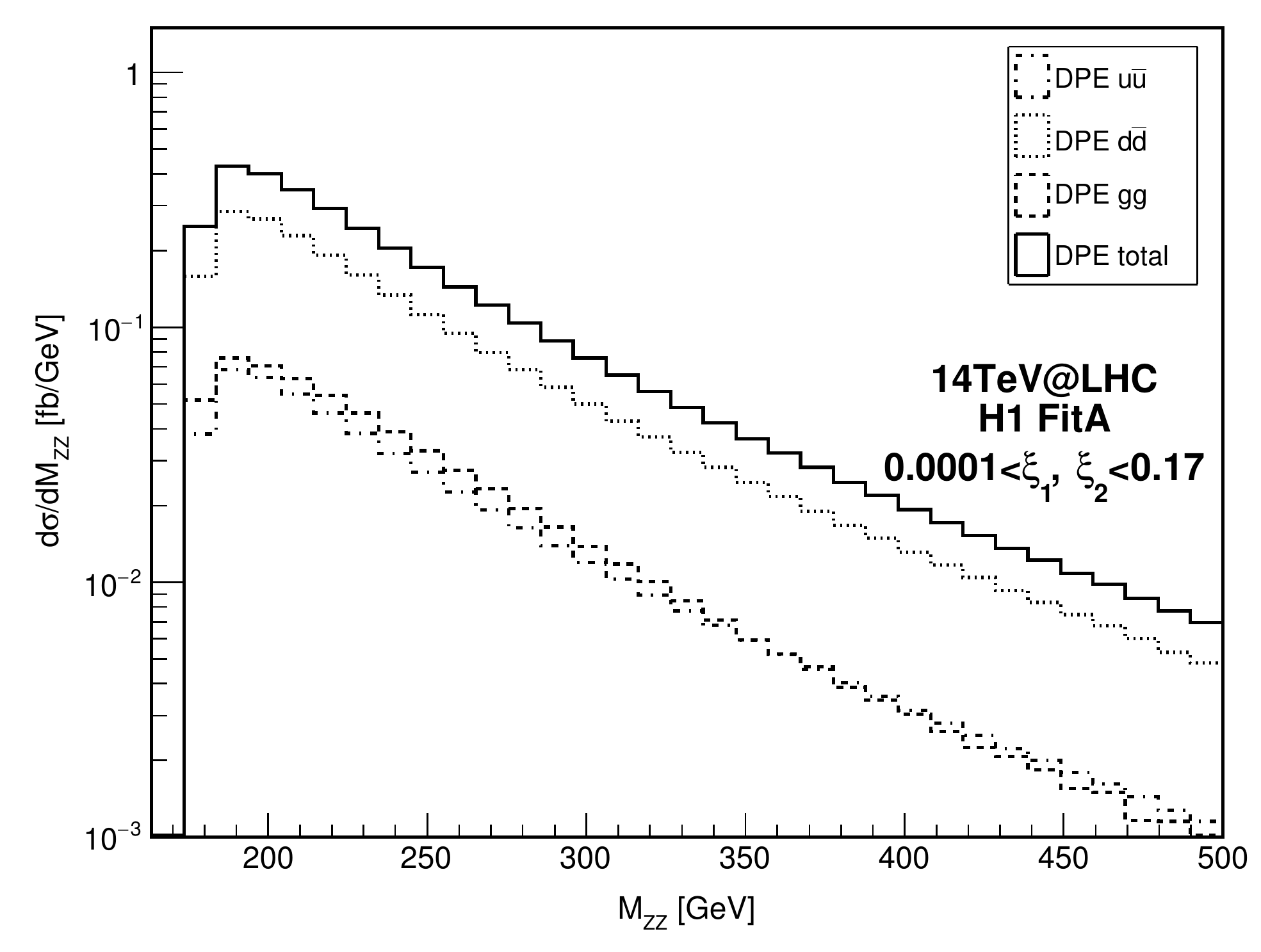}
\caption{\label{14SDx1_mzz}
The invariant mass distribution for the diffractive Z boson pair production at the 14 TeV LHC.
Here we use ``H1 2006 dPDF Fit A''. $0.0001<\xi<0.17$ for TOTEM-CMS is considered. Absorption effects are not included here. }
\end{figure}
In Fig.\ref{14SDx1_mzz} we show the invariant mass distributions of the diffractive Z boson pair production at the 14 TeV LHC.
We compare contributions of single diffractive (first panel) and double Pomeron exchange processes (second panel).
The SD distributions are larger than that of the DPE production by a factor 20 without considering the absorption factor.
We also present the sub-contributions from the up-anti-up quark collision (dash-dotted curve),
down-anti-down quark collision (dotted curve) as well as gluon-gluon fusion (dashed curve).
In any case down-quark collision dominates among the different contributions.
Their sum is plot by the solid curve. As we said, the calculation is done assumes Regge factorization.
Absorption corrections can be taken into account by a multiplicative factor
being a probability of a rapidity gap survival (see e.g. Eq.(\ref{S2})).
Such a factor is approximately $\rm \langle|S|^2\rangle=0.03$ for the LHC energy $\rm \sqrt{s}=14\ TeV$.
The diffractive distributions in the figure should be multiplied in addition by these factors.
In order to avoid model dependence the reader can use his/her own number
when comparing different contributions. Here and in the following the absorption effects are not included for simplicity.

\begin{figure}[hbt!]
\centering
\includegraphics[scale=0.40]{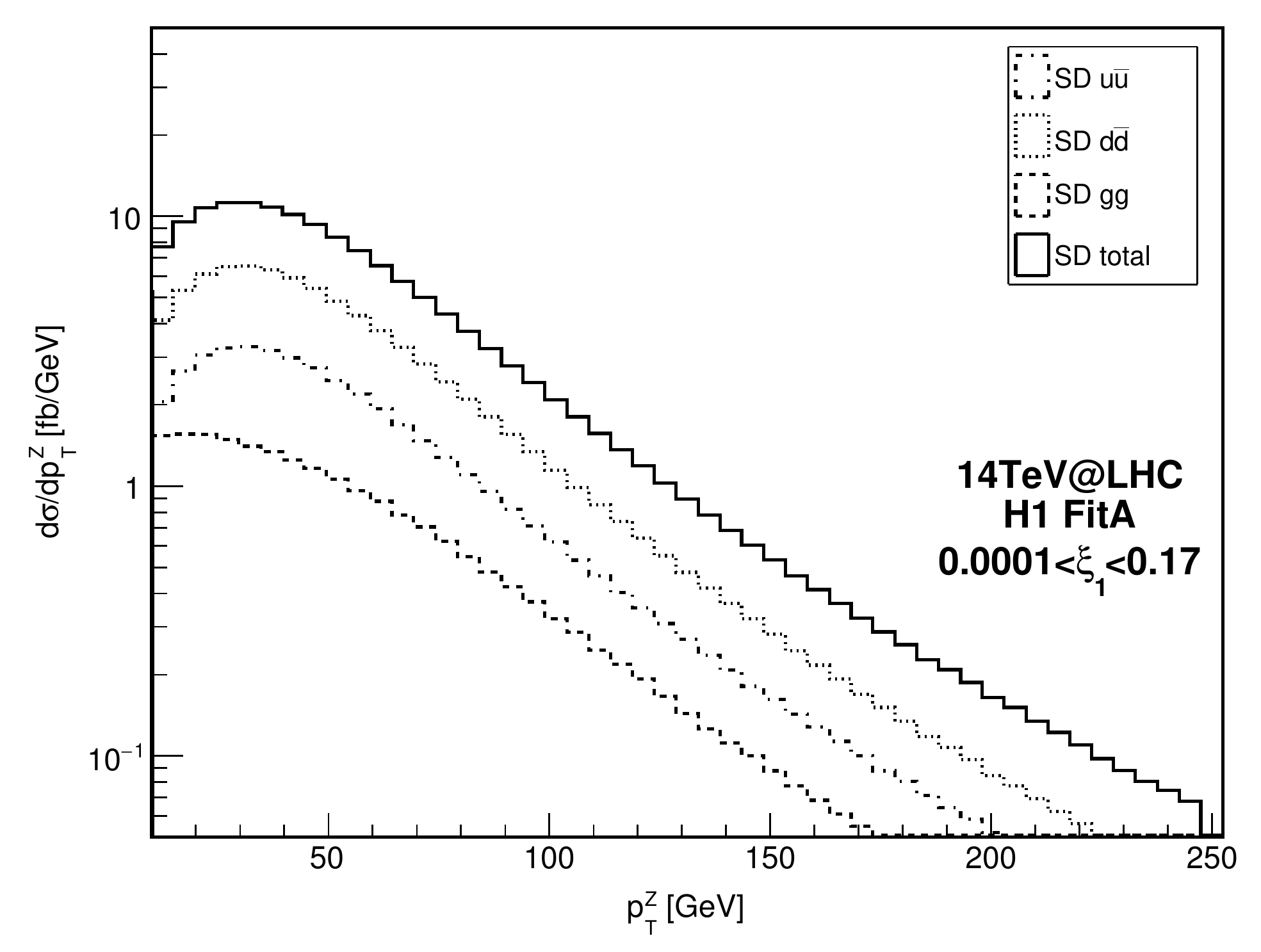}
\includegraphics[scale=0.40]{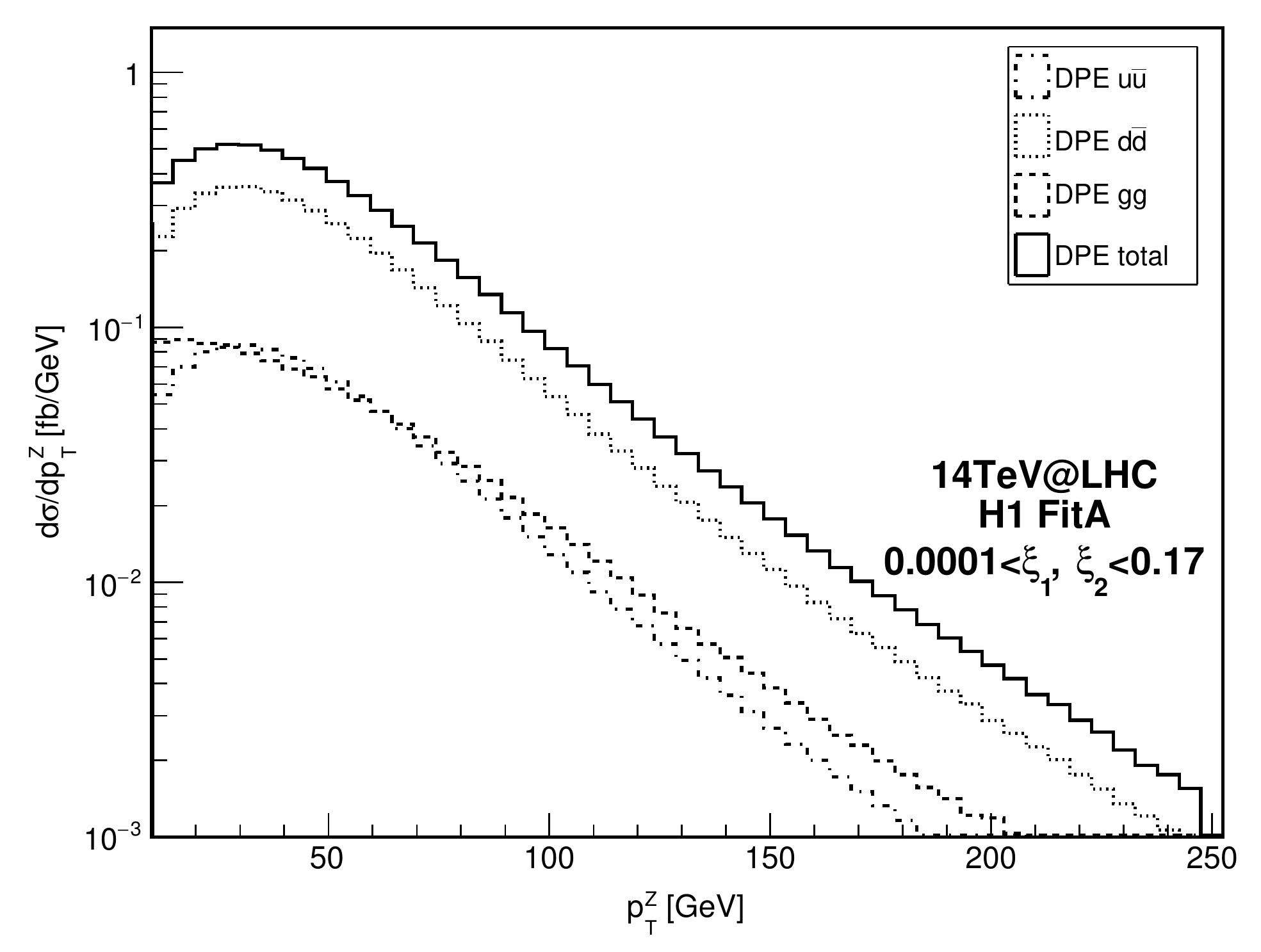}
\includegraphics[scale=0.40]{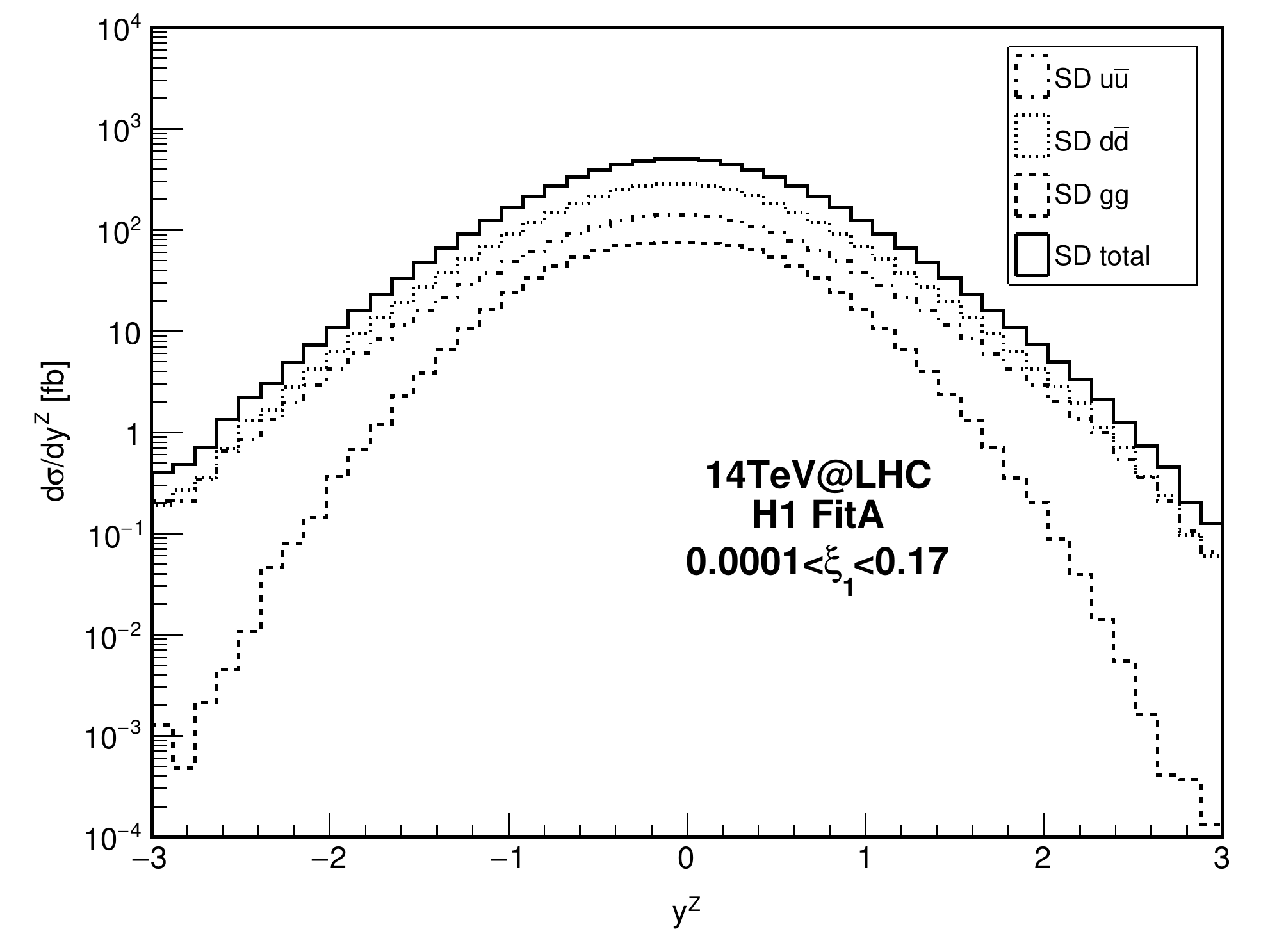}
\includegraphics[scale=0.40]{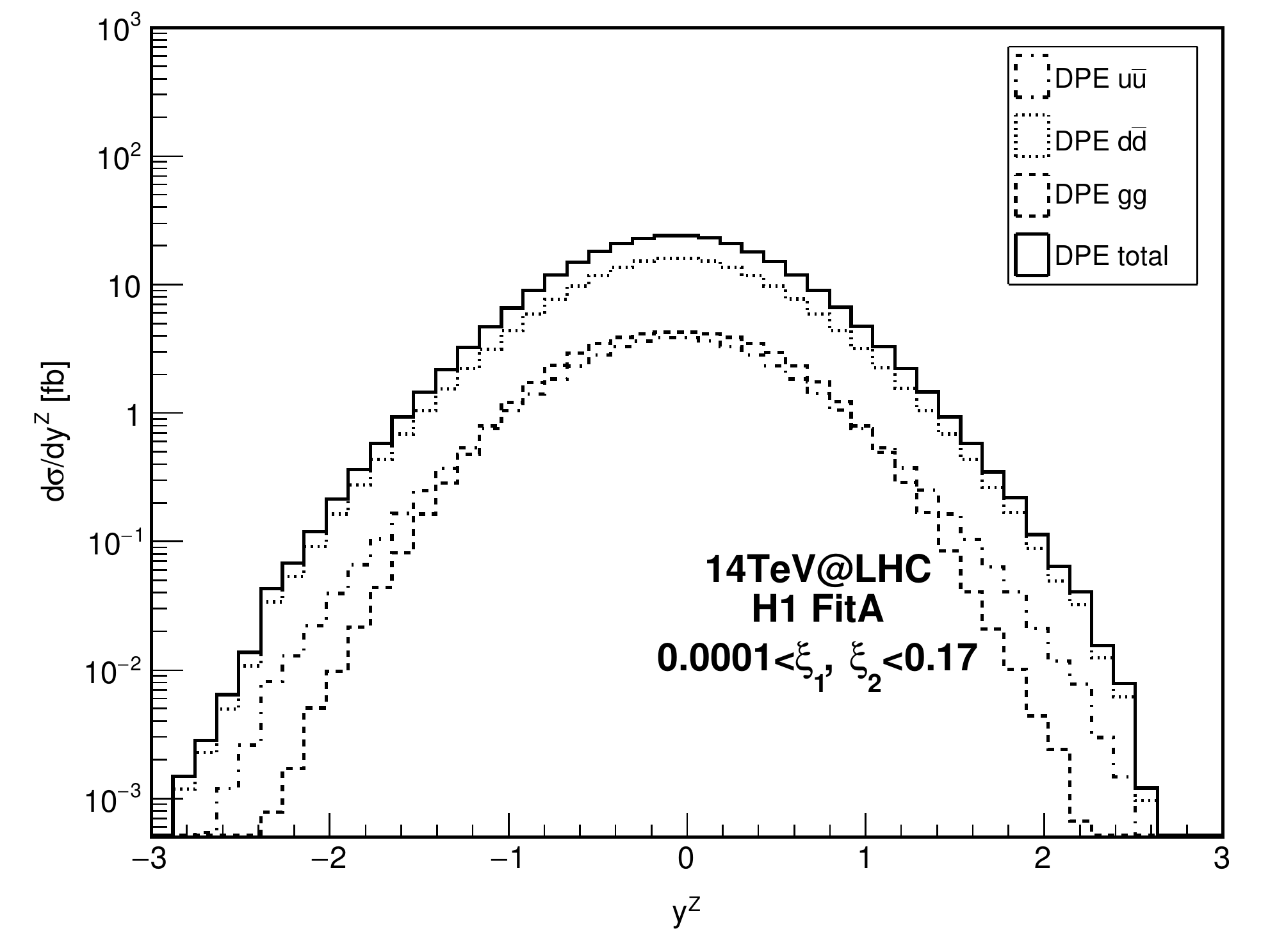}\\
\caption{\label{14SDx1_ptzyz}
The transverse momentum and rapidity distributions for the Z boson at the 14 TeV LHC.
Here we use ``H1 2006 dPDF Fit A''. $0.0001<\xi<0.17$ for TOTEM-CMS is considered. Absorption effects are not included here.}
\end{figure}
In Fig.\ref{14SDx1_ptzyz} we present the Z transverse momentum distribution in the first two panels for SD and DPE production respectively.
As can be seen, its kinematically allowed range extend up to around half of $\rm M_{ZZ}^{max}$.
Given the fast falling nature of the $\rm M_{ZZ}$-distribution, dominated by low values of the invariant,
the Z boson transverse momentum distribution shows a maximum at $\rm p_T\sim M_{ZZ}^{min}/4$.
The rapidity distribution of the Z boson is shown in the second two panels.
Both the SD and DPE contribution as well as sub contributions
are concentrated at mid-rapidities and strongly asymmetric around $\rm y=0$
as a consequence of limiting integration over $\rm x_{\pom}$ in the range $0.0001<\xi<0.17$.

\begin{figure}[hbt!]
\centering
\includegraphics[scale=0.40]{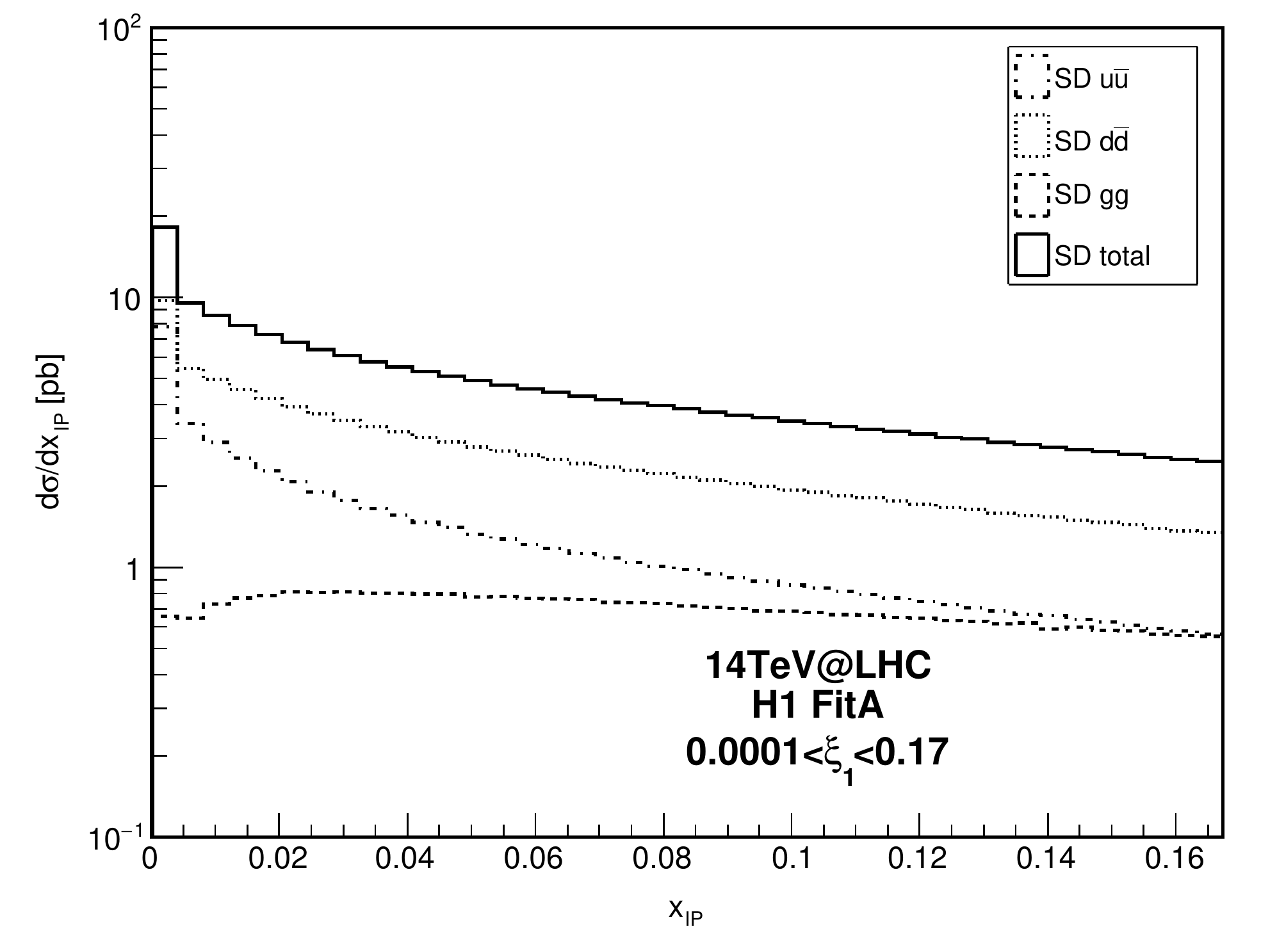}
\caption{\label{14SDx1_xIP}
The $\rm x_{\pom}$ distribution for single diffractive (SD) Z boson pair production at the 14 TeV LHC.
Here we use ``H1 2006 dPDF Fit A''. $0.0001<\xi<0.17$ for TOTEM-CMS is considered. Absorption effects are not included here.}
\end{figure}
In Fig.\ref{14SDx1_xIP} we present the $\rm x_{\pom}$ distribution for single diffractive Z boson pair production.
Still, TOTEM-CMS detector acceptance is considered for simplicity.
We show the up-quark collision, the down-quark collision and the gluon-gluon fusion productions separately and
use solid curve to present their total sum as the function of $\rm x_{\pom}$.
As displayed in the figure, the dominant contribution come from the down-quark
collision which is around two to four times larger than that of the others.
For the up-quark collision and gluon-gluon fusion,
their contributions discrepant largely in the small range of $\rm x_{\pom}$,
while become close to each other as the value of $\rm x_{\pom}$ become larger.
Typically, the quark collision contribution enhance obviously at the small $\rm x_{\pom}$ range,
say, approximately as an inverse power of $\rm x_{\pom}$ at small $\rm x_{\pom}$.
This is not the same as in the gluon-gluon fusion case
where there is some suppression at the small value of $\rm x_{\pom}$.
Nevertheless, the total contribution still show obvious enhancement at small $\rm x_{\pom}$ range.

\begin{figure}[hbt!]
\centering
\includegraphics[scale=0.40]{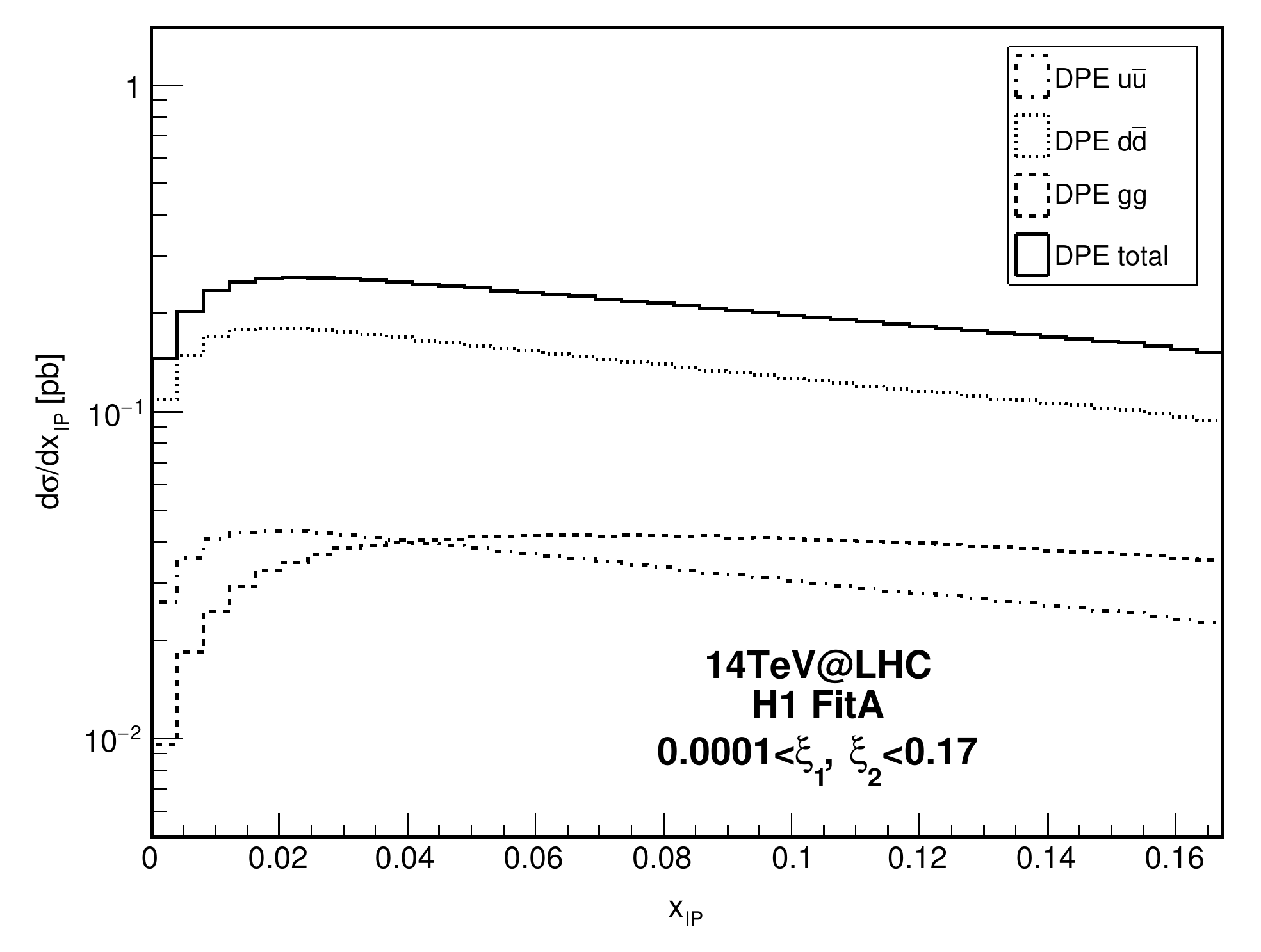}
\includegraphics[scale=0.40]{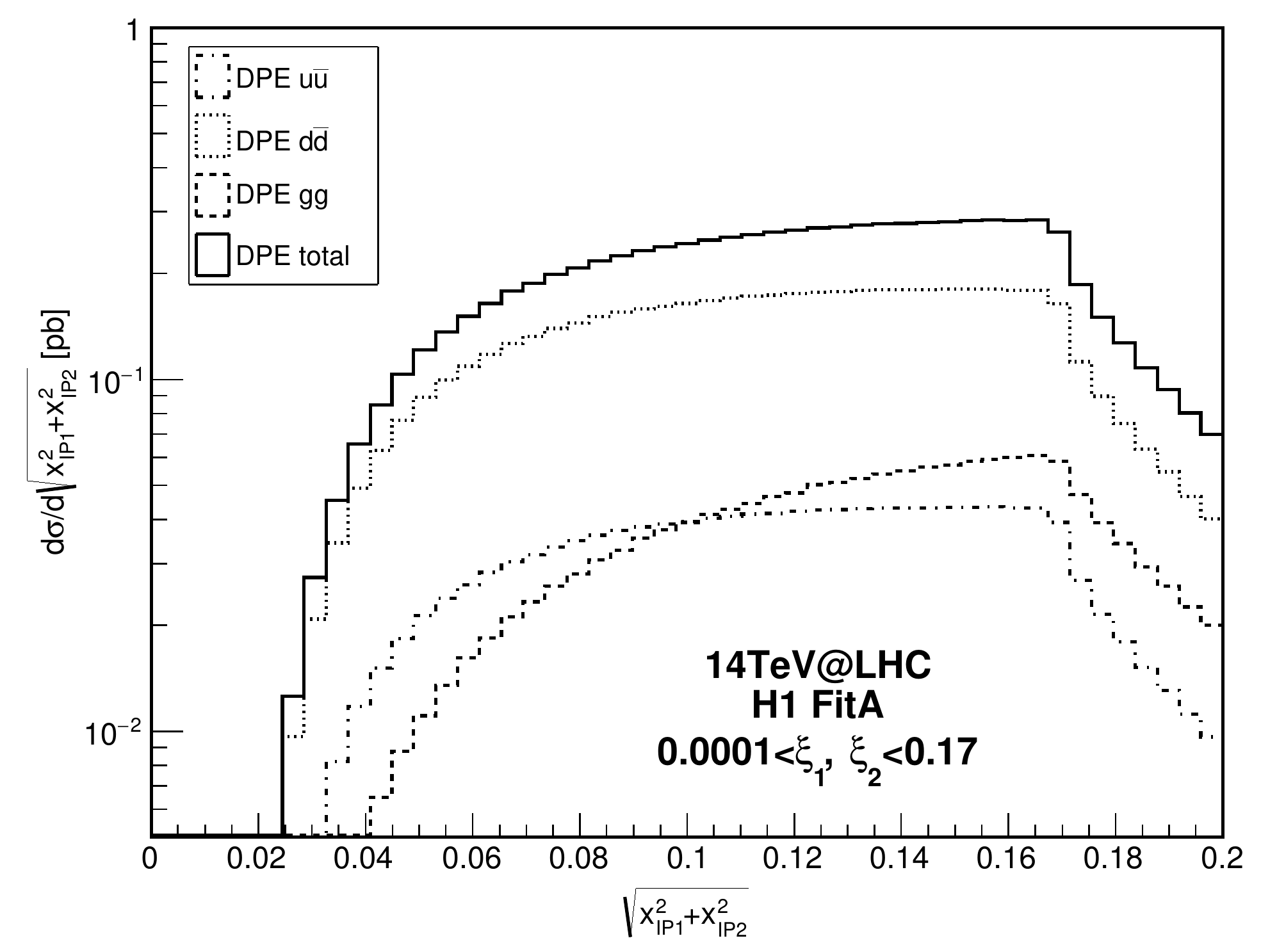}
\caption{\label{14DPEx1_xIP}
The $\rm x_{\pom}$ ($\rm x'_{\pom}$) distribution for double Pomeron exchange (DPE) Z boson pair production at the 14 TeV LHC.
$0.0001<\xi<0.17$ for TOTEM-CMS is considered. Absorption effects are not included here.}
\end{figure}
In order to compare, we display the same distribution in Fig.\ref{14DPEx1_xIP} for double Pomeron exchange Z boson pair production.
$\rm x_{\pom}$ is one fraction of the proton side (first panel).
As can be found in the figure, in contrast with the SD production,
$\rm x_{\pom}$ DPE distribution decreases at small $\rm x_{\pom}$ range
for both the quark-collision and the gluon-gluon fusion.
In order to include the fraction distribution for both sides of the proton in the DPE production,
we define $\rm x'_{\pom}=\sqrt{x^2_{\pom_1}+x^2_{\pom_2}}$ and display its distribution in the second panel in Fig.\ref{14DPEx1_xIP}.
It will be interesting to find out that $\rm x'_{\pom}$ distribution spread mainly in the central range while
on both bound ranges, decreases to small values.
For the front range may due to the large mass of Z boson pair causes that the small value of $\rm x_{\pom_{1,2}}$
are not accessible kinematically, while the behaviour in the ending boundary is due to the forward detector acceptance we considered
that makes a behaviour of the strong suppression.

\begin{figure}[hbt!]
\centering
\includegraphics[scale=0.40]{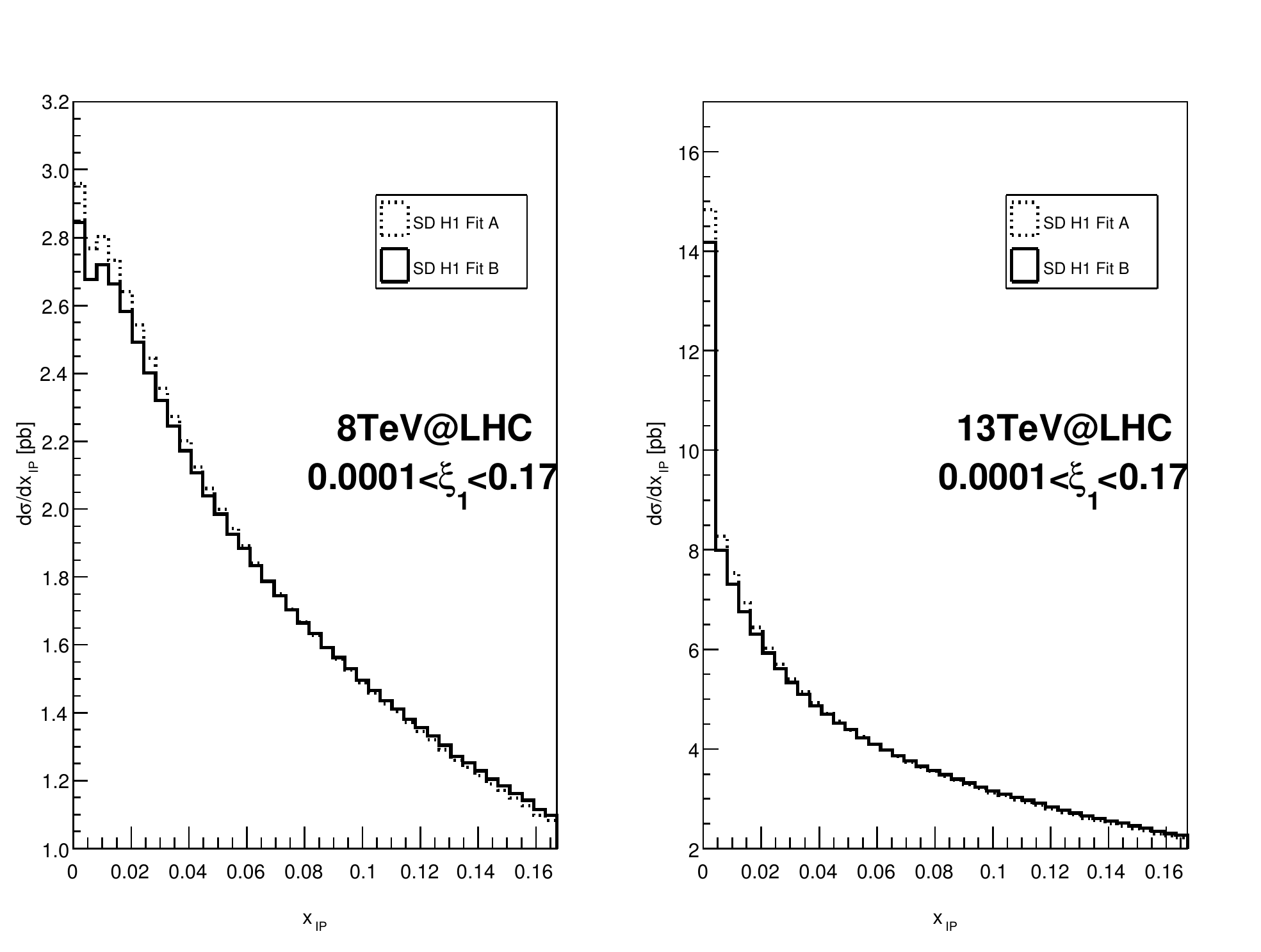}
\includegraphics[scale=0.40]{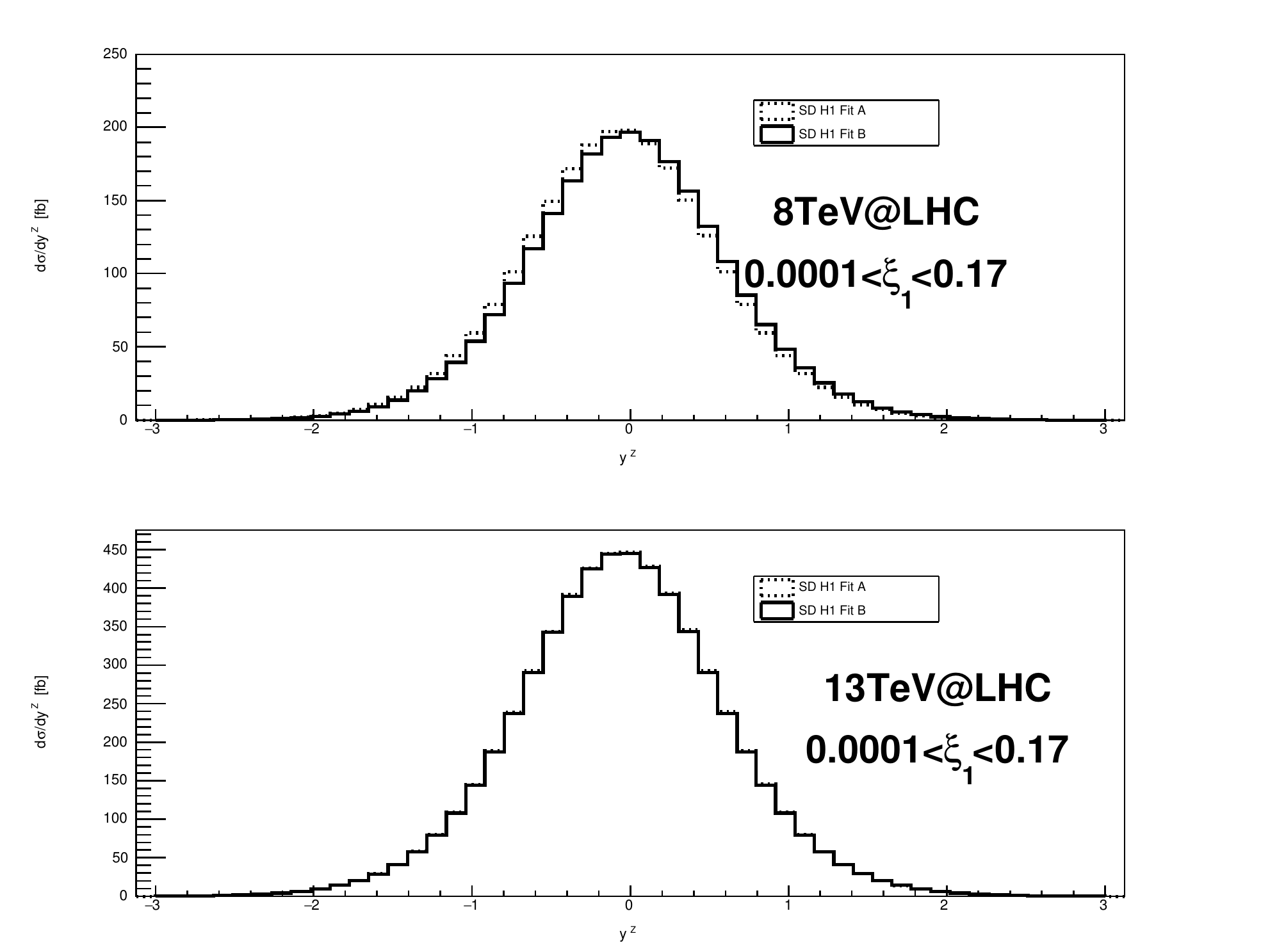}
\caption{\label{SD_FitAB}
Single diffractive (SD) production of $\rm x_{\pom}$ and rapidity distributions at the 8 and 13 TeV LHC
with the use of ``H1 2006 dPDF Fit A'' (solid curves) and ``H1 2006 dPDF Fit B'' (dotted curves).
The detector acceptances are fixed in the range of $0.0001<\xi_1<0.17$. Absorption effects are not included here.}
\end{figure}
\begin{figure}[hbt!]
\centering
\includegraphics[scale=0.40]{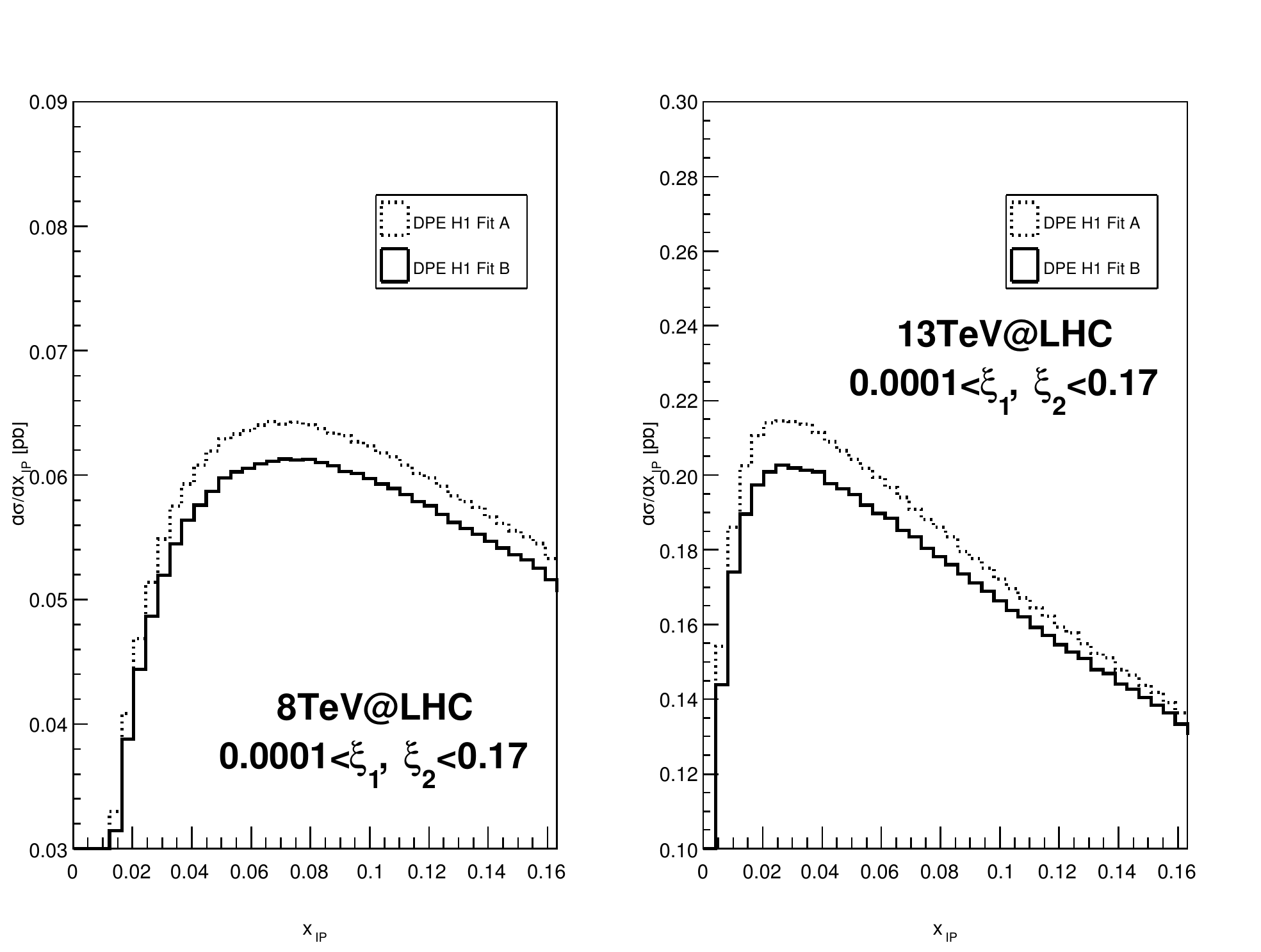}
\includegraphics[scale=0.40]{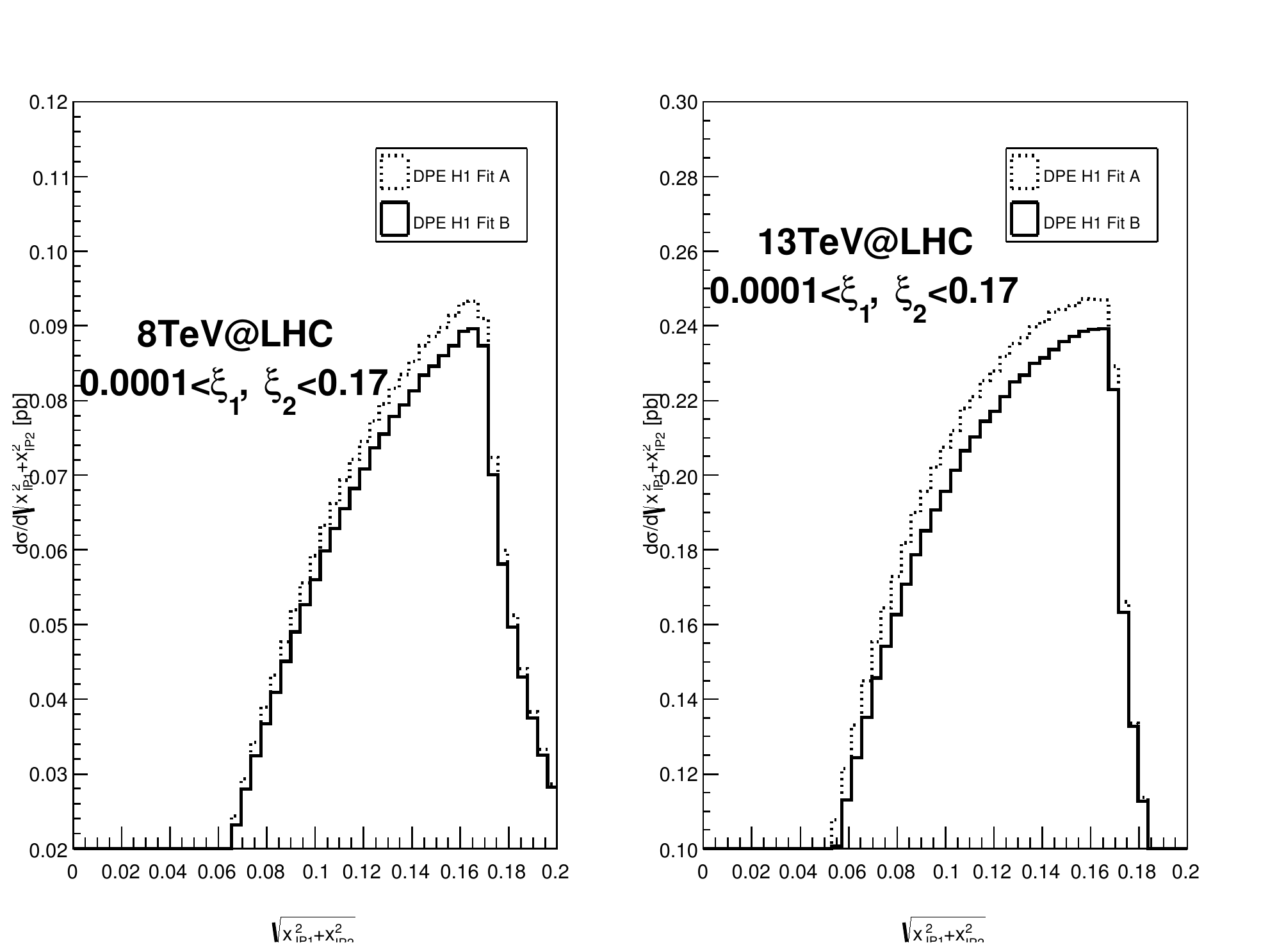}
\includegraphics[scale=0.40]{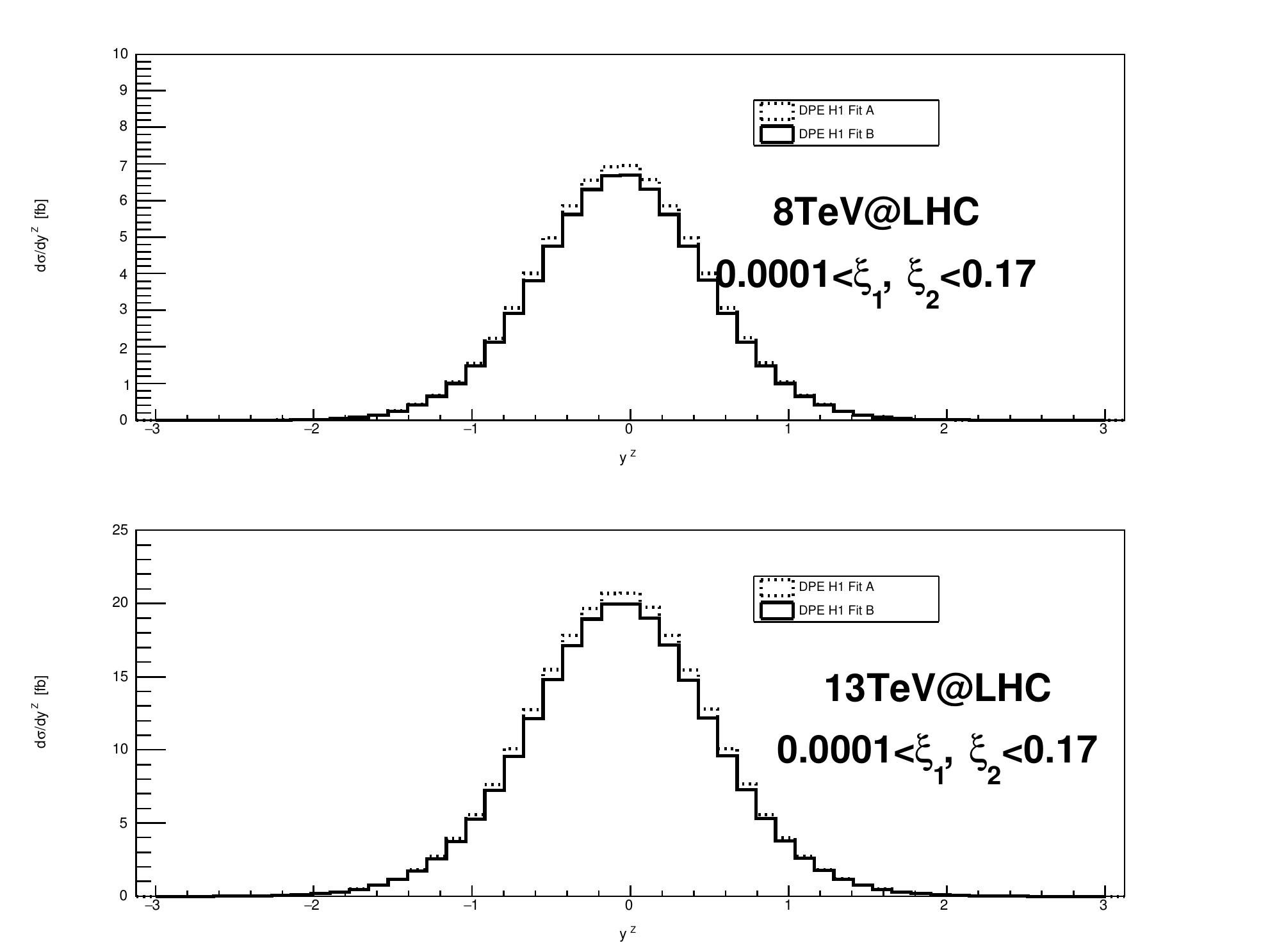}
\caption{\label{DPE_FitAB}
Double Pomeron exchange (DPE) production of $\rm x_{\pom}$, $\rm x'_{\pom}$ and rapidity distributions at the 8 and 13 TeV LHC
with the use of ``H1 2006 dPDF Fit A'' (solid curves) and ``H1 2006 dPDF Fit B'' (dotted curves).
The detector acceptances are fixed in the range of $0.0001<\xi_1<0.17$. Absorption effects are not included here.}
\end{figure}
The first uncertainty in diffractive productions is the gap survival probability as we mentioned above.
Another error represents the propagation of experimental uncertainties is obtained in the diffractive PDF fit.
We shown this in Fig.\ref{SD_FitAB} and Fig.\ref{DPE_FitAB} for SD and DPE production.
Results of 8, 13 TeV and distributions of $\rm x_{\pom}$ and rapidity are presented as examples.
The detector acceptance is fixed in the range of $0.0001<\xi_1<0.17$ where similar results
can be obtained for $0.015<\xi_1<0.15$. The ``H1 2006 dPDF Fit A'' (solid curves) is considered,
whereas a replacement by ``H1 2006 dPDF Fit B'' (dotted curves) keeps the results
slightly different. For the PDFs in the proton we have always considered the cteq6L1 parameterization\cite{CTEQ6}.
As can be seen, the discrepancy induced by using different fits in DPE production is a little larger than that in SD production.
For all $\rm x_{\pom}$, $\rm \sqrt{x^2_{\pom_{1}}+x^2_{\pom_{2}}}$ and rapidity distributions,
the small enhancement showed mainly at the peak range. Nevertheless,
there is no large discrepancy observed, therefore, the uncertainty is small
in using the different fit procedures for diffractive PDFs.

\begin{figure}[hbt!]
\centering
\includegraphics[scale=0.40]{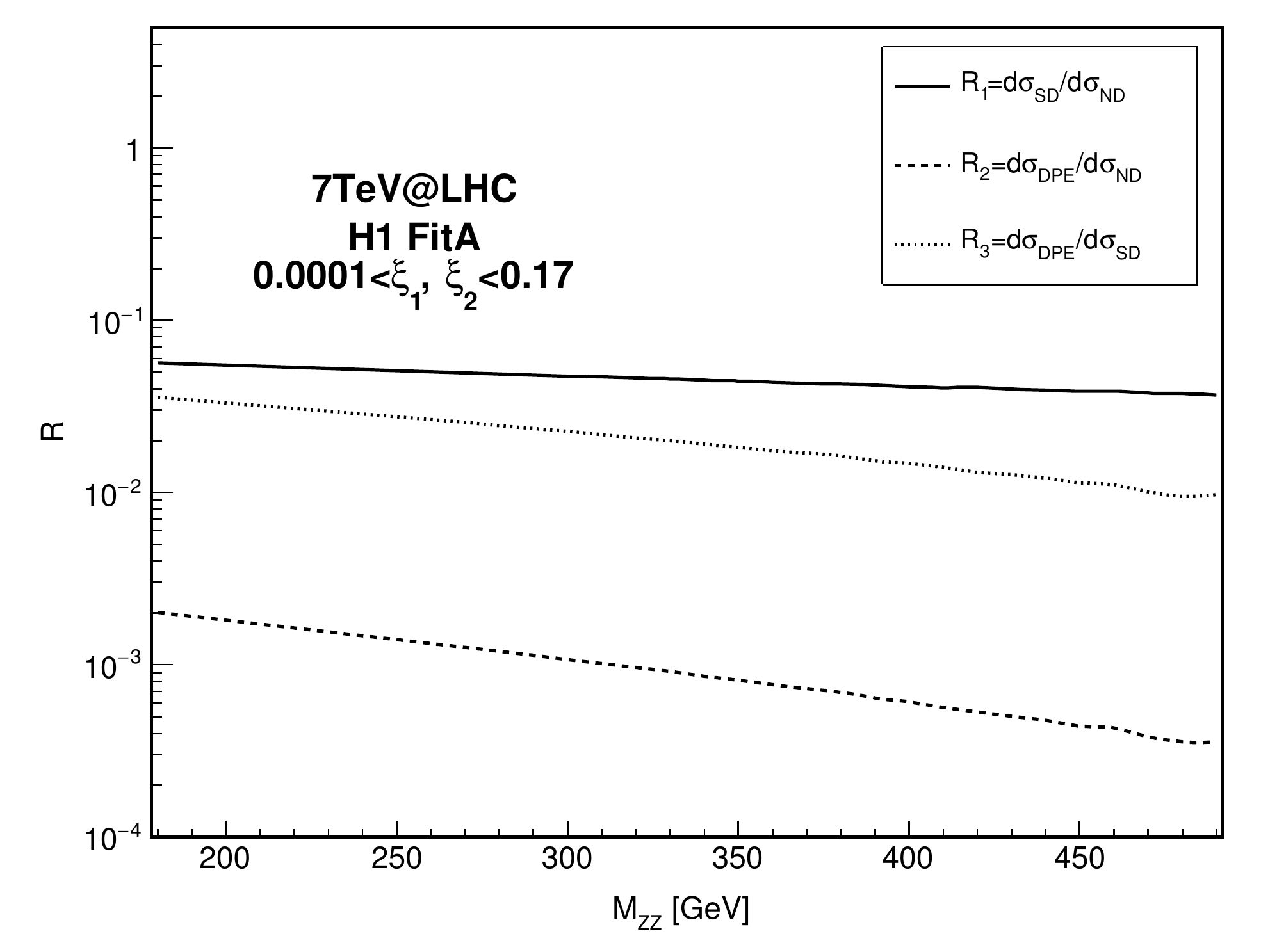}
\includegraphics[scale=0.40]{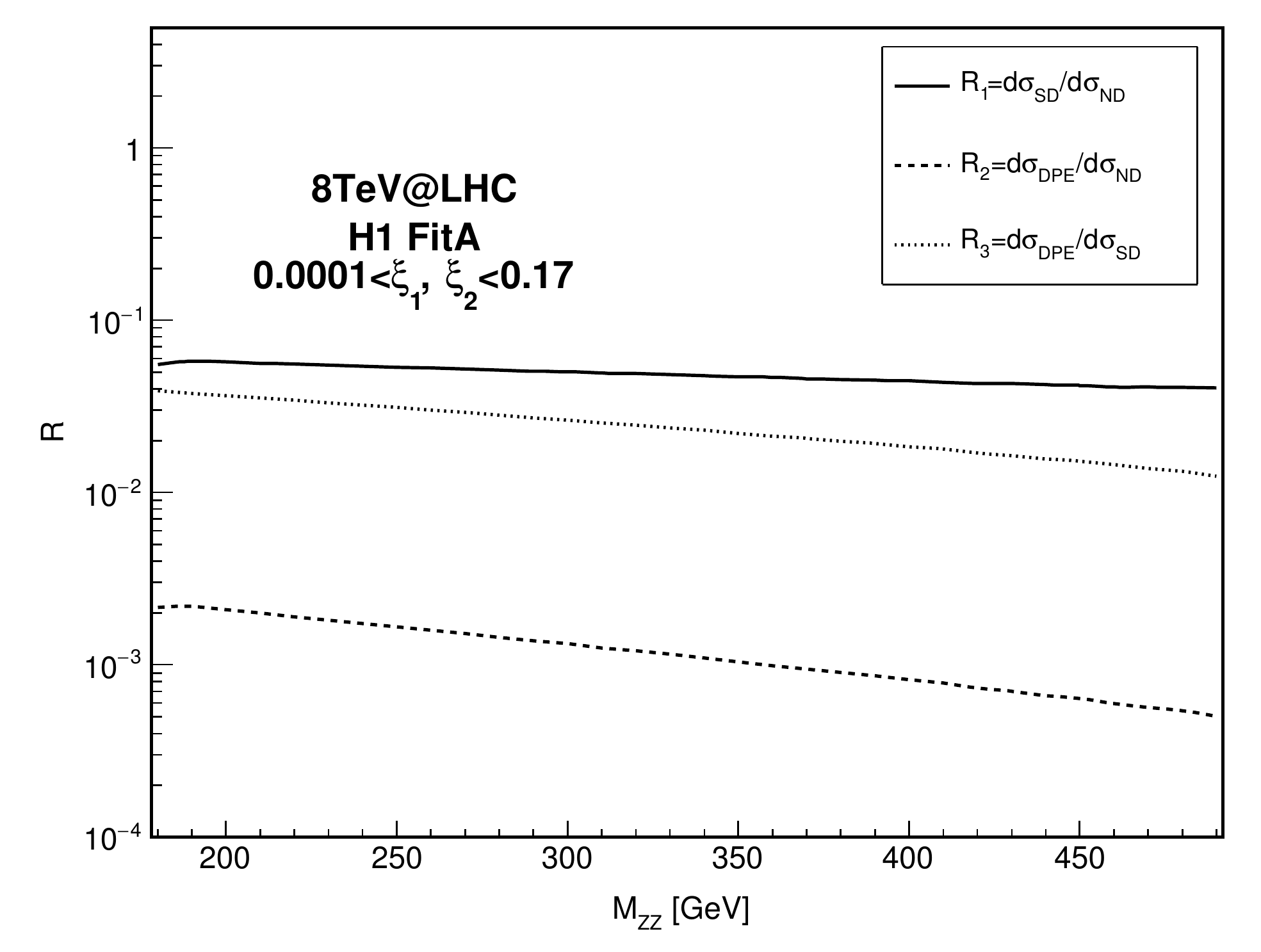}
\includegraphics[scale=0.40]{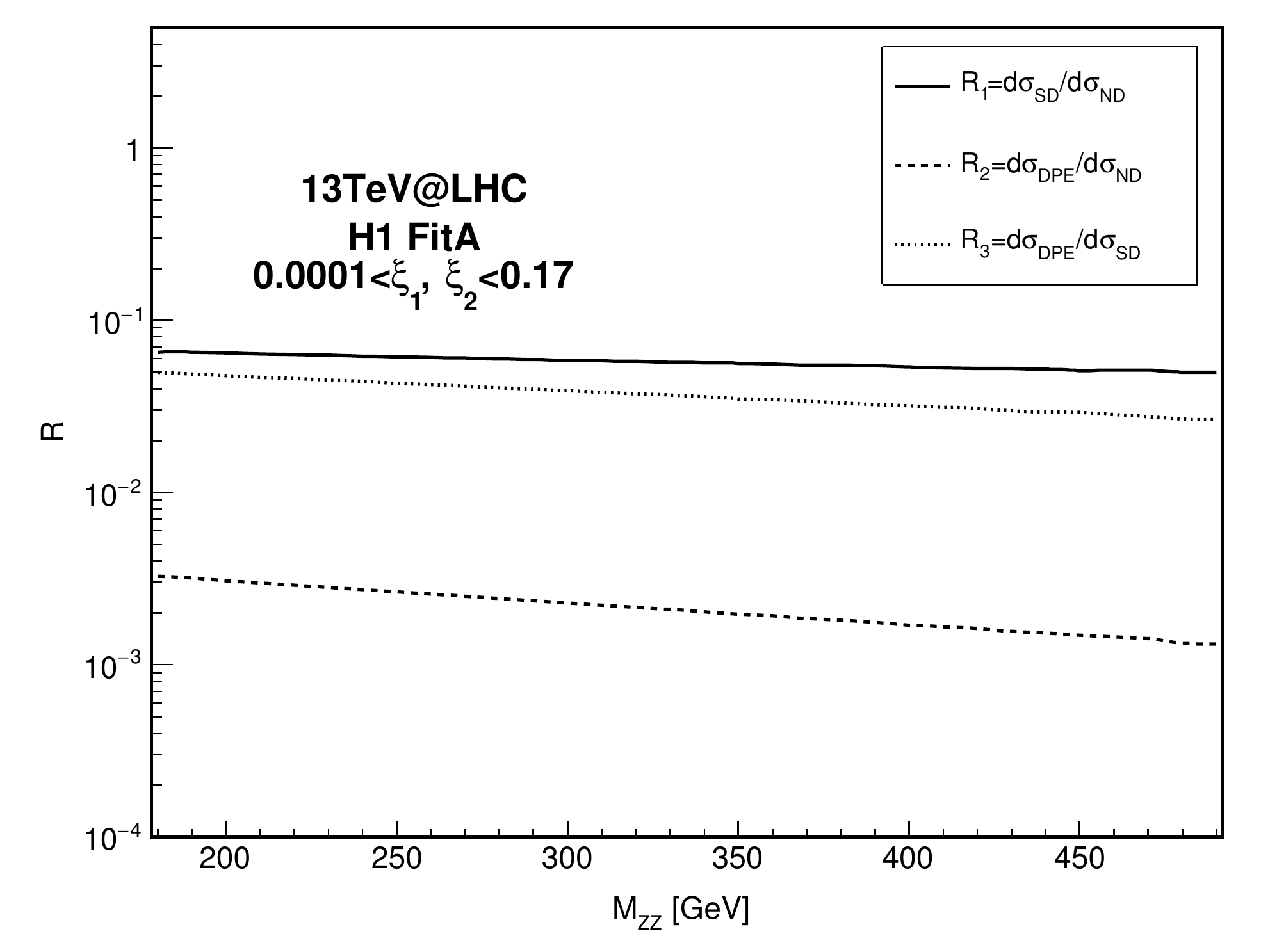}
\includegraphics[scale=0.40]{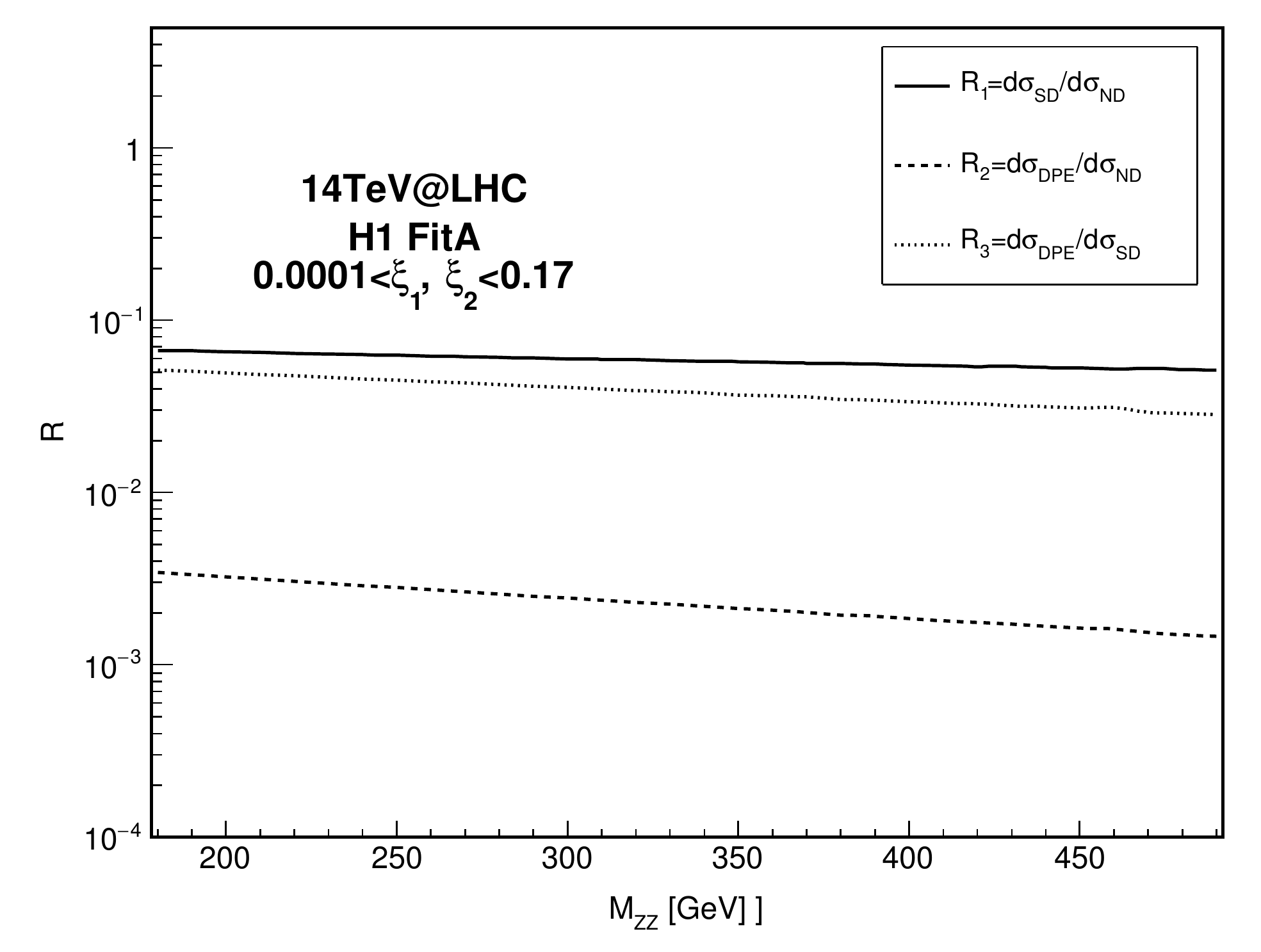}
\caption{\label{Ratio}
R ratios as a function of the invariant mass $\rm M_{ZZ}$ for different values of the LHC energy.
$0.0001<\xi<0.17$ for TOTEM-CMS is considered. Absorption effects are not included here.}
\end{figure}
The third uncertainty, of theoretical nature, is obtained by varying the factorisation scales.
Such uncertainties can be reduced by including higher order corrections
whereas the complete calculation is out the scope here.
In the present content, we stable against factorisation scale variation conveniently
by considering appropriate ratios of diffractive over non-diffractive (ND) cross sections
\begin{eqnarray}
\rm R=\frac{\sigma(pp\to pYX)}{\sigma(pp \to YX)}\ \ and \ \ R=\frac{\sigma(pp\to pYXp)}{\sigma(pp \to YX)},
\end{eqnarray}
or DPE cross section over the SD ones
\begin{eqnarray}
\rm R=\frac{\sigma(pp\to pYXp)}{\sigma(pp \to pYX)},
\end{eqnarray}
which also offer the advantage to reduce experimental systematics errors.
Here Y stands for the selected hard scattering process (Z boson pair this case)
and X for the unobserved part of the final states.
At the Tevatron the ratio R has been measured in a variety of
final states \cite{R_Tevatron_Bj}\cite{R_Tevatron_WZ}\cite{R_Tevatron_jj}
and show some stable behaviour with a value close to one percent.
Typically, in our case considering at the distribution level,
we define the single diffractive ratio as
\begin{eqnarray}
\rm R_{1}=\frac{d\sigma_{SD}}{d\sigma_{ND}},
\end{eqnarray}
the double Permon diffractive ratio by
\begin{eqnarray}
\rm R_{2}=\frac{d\sigma_{DPE}}{d\sigma_{ND}},
\end{eqnarray}
and also DPE over SD ratio as
\begin{eqnarray}
\rm R_{3}=\frac{d\sigma_{DPE}}{d\sigma_{SD}}.
\end{eqnarray}
As predicted in Fig.\ref{Ratio}, we plot the R ratio as a function of $\rm M_{ZZ}$ distribution
with solid curve for $\rm R_1$, dashed curve for $\rm R_2$ and dotted curve for $\rm R_3$, respectively.
Based on these results we verify that, for the single diffractive Z boson pair production in pp collision,
given leading order estimate of the non-diffractive cross section, the ratio $\rm R_1$ is varies
between $5\%$ and $7\%$ and decreases mildly as a function of the invariant mass of the Z boson pair.
The double Pomeron exchange productions are about 20-100 times smaller than that of the single diffractive ones,
as can be found in the DPE over SD ratio $\rm R_3$, varies between $1\%-4\% (3\%-5\%)$ for 7,8 (13, 14) TeV correspondingly.
For the double Pomeron exchange, the ratio $\rm R_2$ varies in the range $0.03\%-0.2\%(0.1\%-0.3\%)$
for 7,8 (13, 14) TeV, which are much smaller than that of $\rm R_1$.
By the definition of R parameters, predictions affected by large theoretical errors
associated with scale variations can be reduced in a simple way.
These predictions however does not take into account the gap survival suppression factor.
With this respect it would be still interesting to check whether the data follow at least the shape
of the ratio as a function of $\rm M_{zz}$ as we shown in the future measurements.

\begin{figure}[hbtp]
\centering
\includegraphics[scale=0.6]{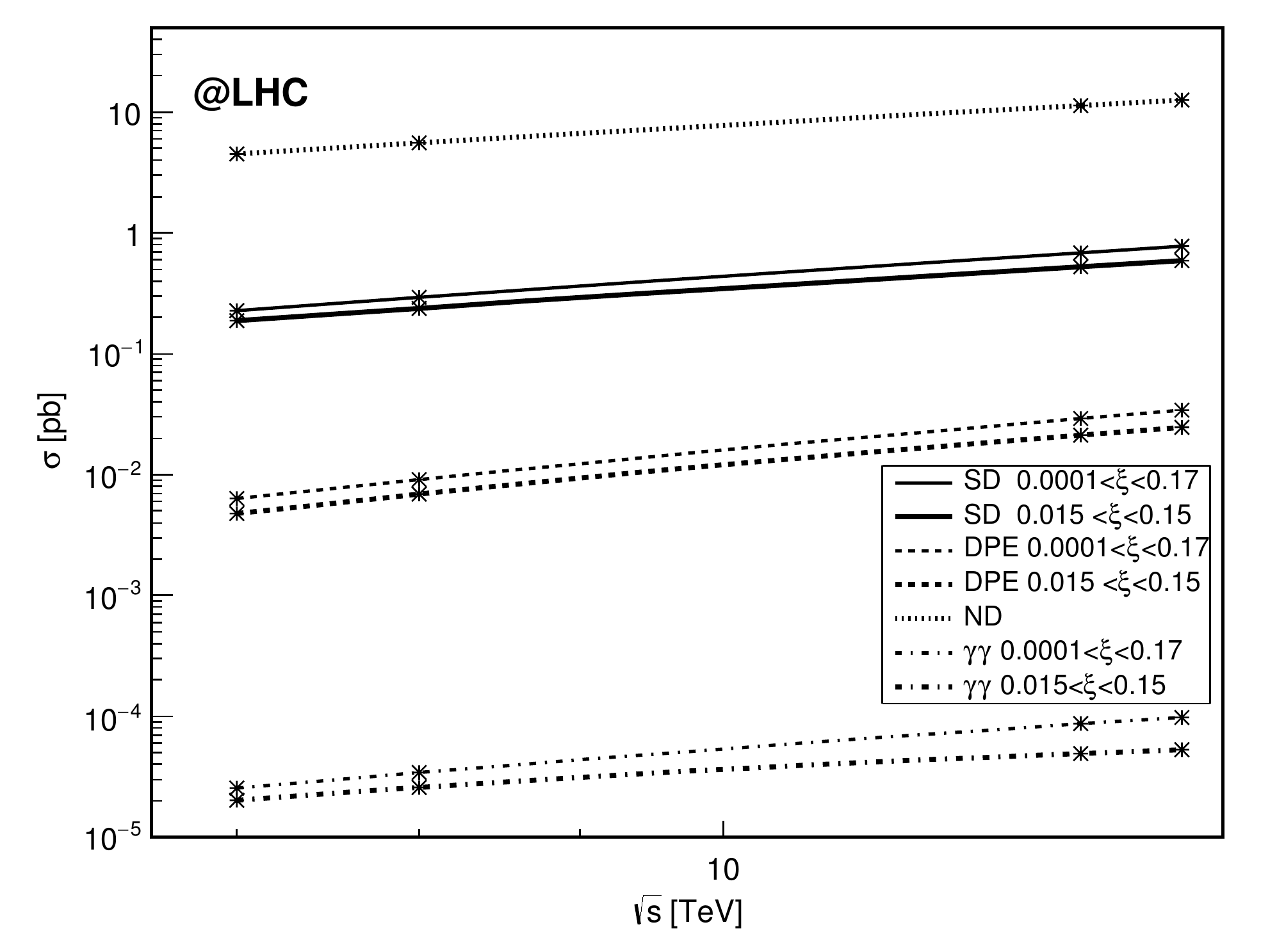}
\caption{\label{Xsection}
The total cross sections (in unit of pb) for single diffractive (SD), double Pomeron exchange (DPE),
photon-photon induced ($\gamma\gamma$) and non-diffractive (ND) Z boson pair reactions
as a function of proton-proton center-of-mass energy at the 7, 8, 13 and 14 TeV LHC.
Both the detector acceptances of $0.0001<\xi<0.17$ and $0.015<\xi<0.15$ are considered.
The rapidity gap survival probability factor is not taken into account here.}
\end{figure}
Finally in Fig.\ref{Xsection} we show the total cross sections (in unit of pb)
for the single diffractive (SD) and double Pomeron exchange (DPE) cross sections,
and compare to the photon-photon induced ($\gamma\gamma$) as well as non-diffractive (ND) Z boson pair reactions,
as a function of proton-proton CMS energy of 7, 8, 13 and 14 TeV at the LHC.
We use solid, dashed, dotted and dash-dotted curves to present SD, DPE, $\gamma\gamma$ and ND cross sections, respectively.
For SD, DPE and $\gamma\gamma$ production, both the detector acceptances of
$0.0001<\xi<0.17$ (thin curve) and $0.015<\xi<0.15$ (thick curve) are considered.
Notice here the rapidity gap survival probability factor is not taken into account.
Features can been found in the figures are list as the following:
\begin{itemize}
 \item The cross sections for different production mechanisms increase linearly as functions of the colliding energy.
 \item Typical size order is normally $\rm \sigma_{ND}>\sigma_{SD}>\sigma_{DPE}>\sigma_{\gamma\gamma}$ as excepted.
 \item Results from considering ATLAS-AFP detector acceptance ($0.015<\xi<0.15$) are comparable with that from TOTEM-CMS ($0.0001<\xi<0.17$) but a little smaller.
\end{itemize}
When the rapidity gap survival probability factor is considered, we can find
that the SD cross section is at the order of $\rm \sim {\cal O}(10\ fb)$.
For the DPE production rate is about $\rm 0.1-1\ fb$ which is small but still larger than that of $\gamma\gamma$ induced production which is only about 0.1 fb.
The smallness of the Z boson pair production, however, is not a thoroughly bad thing.
As we said, when go to LHC energy frontier, exclusive production
may open a new window to new physics searching\cite{pp2pZZp_Gupta}\cite{pp2pZZp_Chapon} while in this case diffractive may serve as the important background.
If a new sector is produced through gauge Z boson pair production, such mechanism can be tested with a typical clean environment.

\section{CONCLUSION}

A rich program at the Large Hadron Collider (LHC) is being pursued
in diffractive physics by all collaborations either based on the identification
of large rapidity gaps or by using dedicated proton spectrometers.
In our present study, we perform the calculation from phenomenological analysis of Z boson pair
hard diffractive production at the LHC. Our calculation is based on the Regge factorization approach.
Diffractive parton density functions (dPDFs) extracted by the H1 Collaboration at DESY-HERA are used
and uncertainties by using different fits in the dPDFs are discussed.
The multiple Pomeron exchange corrections are considered through the rapidity gap survival probability factor.
We display various kinematical distributions for both the single diffractive(SD) and double Pomeron exchange (DPE) productions.
We give also numerical predictions for their cross sections and compare with the photon-induced and non-diffractive ones.
The contributions from both quark-anti-quark collision and gluon-gluon fusion modes are displayed and compared.
We define the appropriate ratios of diffractive over non-diffractive (ND) productions,
by using which predictions affected by theoretical errors associated with scale variations can be reduced.
Typically the single diffractive ratio is varies between $5\%$ and $7\%$
while the double Pomeron exchange ratio varies in the range $0.03\%-0.2\%(0.1\%-0.3\%)$ for 7,8 (13, 14) TeV.
We make predictions which could be compared to future measurements at the LHC
where forward proton detectors are installed and detector acceptances are considered.

\section*{Acknowledgments} \hspace{5mm}
This work is supported by the National Natural Science Foundation of China
(Grant No. 11675033), by the Fundamental Research Funds for the Central Universities
(Grant No. DUT15LK22).

\end{document}